%% file: ANA-HMBS-2024-13-PAPER.tex
\pdfoutput=1
\documentclass[PAPER, atlasdraft=false, cernpreprint, UKenglish, texmf, orcidlogo]{atlasdoc}
\usepackage{atlaspackage}
\usepackage{multirow}

\usepackage{atlasbiblatex}

\usepackage[process=true,BSM=true]{atlasphysics}
\graphicspath{{logos/}{figures/}}

\usepackage{ANA-HMBS-2024-13-PAPER-defs}

\AtlasTitle{A search for dark matter produced in association with a dark Higgs boson decaying into a Higgs boson pair in 3$b$ or 4$b$ final states using $pp$ collisions at $\sqrt{s}=13$ TeV with the ATLAS detector}

\AtlasAbstract{%
A search is performed for dark matter particles produced in association with a resonant pair of Higgs bosons using $140$ fb$^{-1}$ of proton--proton collisions at a centre-of-mass energy of $13$ TeV recorded by the ATLAS detector at the Large Hadron Collider. This signature is expected in some extensions of the Standard Model predicting the production of dark matter particles, and is interpreted in terms of a dark Higgs model containing a $Z'$ mediator in which the dark Higgs boson $s$ decays into a pair of Higgs bosons. The dark Higgs boson is reconstructed through final states with at least three $b$-tagged jets, produced by the pair of Higgs boson decays, in events with significant missing transverse momentum consistent with the presence of dark matter.
The observed data are found to be in good agreement with Standard Model predictions, constraining scenarios with dark Higgs boson masses within the range of 250 to 400 GeV and $Z'$ mediators up to 2.3 TeV.
}

\AtlasRefCode{HMBS-2024-13}

\PreprintIdNumber{CERN-EP-2025-141}

\AtlasJournal{JHEP}
\AtlasJournalRef{JHEP 09 (2025) 067}
\AtlasDOI{https://doi.org/10.1007/JHEP09(2025)067}

\AtlasCoverEgroupAnalysisTeam{atlas-hmbs-2024-13-analysis-team@cern.ch}

\hypersetup{pdftitle={ATLAS document},pdfauthor={The ATLAS Collaboration}}

\begin{document}

\maketitle

\tableofcontents

\renewcommand{\arraystretch}{1.2}

\section{Introduction}
\label{sec:intro}


%
The existence of dark matter (DM) is widely supported by astrophysical evidence~\cite{Trimble1987,Bertone2005,Feng2010,Porter:2011nv} including galactic rotation velocities of stars~\cite{Corbelli:1999af,Rubin:1980,Begemann:1991}, gravitational lensing~\cite{Clowe:2006eq,Randall:2008ppe,Bradac:2008eu} and precise measurements of the cosmic microwave background~\cite{Hinshaw:2012aka,Akrami:2018vks}.
There is no suitable candidate in the Standard Model (SM), thus its particle nature and interactions remain key open questions in particle physics, motivating many beyond-the-SM (BSM) scenarios.

Complementary approaches are seeking evidence of dark matter particles.
These include searches for direct detection of DM scattering from electrons or nuclei~\cite{LZ:2024zvo,LZ:2022lsv,XENON:2024znc,XENON:2023cxc,PandaX:2024qfu,DarkSide-50:2022qzh}; indirect searches for the product of DM annihilation or decay~\cite{Acharyya:2023ptu,HESS:2022ygk,CTA:2020qlo,Ackermann_2015,TheFermi-LAT:2017vmf,PerezdelosHeros:2020qyt}; gravitational, astrophysical and cosmological probes~\cite{LIGOScientific:2021ffg,Vermeulen:2021epa}; and searches for evidence of dark matter production at colliders~\cite{EXOT-2018-62-fixed,EXOT-2018-64,CMS-EXO-23-005} with particle detectors such as the ATLAS experiment~\cite{PERF-2007-01} at the Large Hadron Collider (LHC)~\cite{Evans:2008zzb}.

Collider particle physics experiments have advantages in this endeavour --- they are sensitive across a wide range of types and masses of proposed DM candidates, and have the potential to characterise its interactions, making them complementary to other types of searches.
One class of DM candidates is stable, electrically neutral, weakly interacting massive particles (WIMPs), motivated by the observation that a candidate with interactions and mass of the order of the weak scale would reproduce the observed relic density~\cite{Planck:2018vyg} via the freeze-out mechanism: a paradigm known as the \textit{WIMP miracle}~\cite{Steigman:1984ac}.
Consequently colliders could be a natural place to find evidence for WIMP DM under well-controlled experimental conditions.

At colliders, it is assumed that dark matter interacts via \textit{mediator} particles, which interact with both DM and SM particles.
New mediator particles can be searched for directly, via their decay into SM particles, or may decay into DM\@.
DM produced at colliders would not interact with the detector, and its presence must be inferred by an imbalance in the total momentum in the plane perpendicular to the collision axis.
This quantity is the missing transverse momentum, with magnitude denoted \ptmiss (or \MET).
Some visible radiation \(X\) must also be present to create this imbalance, creating what is known as an \(X+\ptmiss\) signature.

This paper presents a search for dark matter produced in association with a new scalar mediator, which decays into a resonant pair of SM Higgs bosons. The decay of each Higgs boson (\(h\)) into
two \(b\)-quark-initiated jets, results in a resonant \(hh(\rightarrow b\bar{b}b\bar{b})+\ptmiss\) signature.
This signature, of large \ptmiss\ consistent with the presence of DM and two Higgs bosons, is sensitive to many DM models with extended Higgs sectors or heavy (pseudo-)scalar mediators~\cite{Blanke_2019,Argyropoulos:2024yxo}.

To optimise and interpret the results of the search, a \textit{dark Higgs} model is used~\cite{Duerr:2017uap}.
Motivated by the Higgs mechanism for electroweak symmetry breaking in the SM, this model explains massive fermionic dark matter (\(\chi\)) through a similar mass generation mechanism in the dark sector, positing a \textit{dark Higgs} boson, (\(s\)).
Such a weak-scale dark Higgs boson, interacting through mixing with the SM Higgs boson, will decay analogously to an SM Higgs boson of that mass.
Thus the primary decays are into a pair of \(b\)-quarks at lower masses (\(\ms < \SI{150}{\GeV}\)), with diboson decays (\(WW, ZZ, hh\)) dominating at higher masses. Decays of \(s\) into a top quark pair can occur at low rate, whilst decays into the dark matter particles may be dominant, if kinematically allowed.
An independent motivation for these models is an improved ability to reproduce the observed relic density, due to the presence of additional annihilation channels featuring the dark Higgs boson, such as \(\chi\chi \rightarrow s s \), with subsequent decay of \(s\) into SM particles.
This avoids the over-population common in simple or simplified models with weak-scale mediators~\cite{EXOT-2018-62-fixed}.
Thus the relic density depends primarily upon the coupling between DM and \(s\), whose SM couplings can be small, and hard to observe with traditional collider DM signatures.

\begin{figure}[t]
\centering
\includegraphics[width=0.5\textwidth]{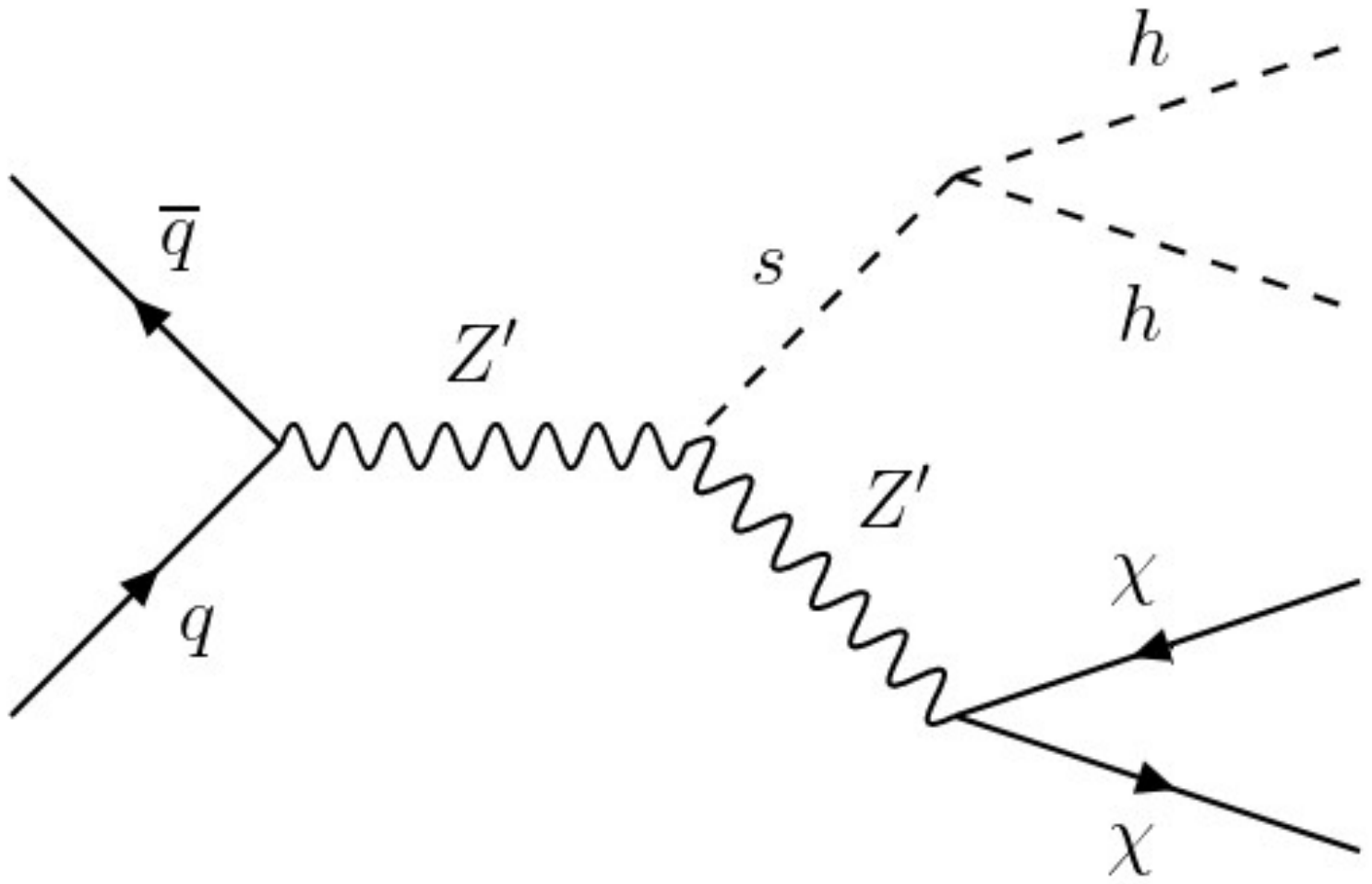}
\caption{Representative signal diagram for resonant \(q\bar q \to Z' \to s\chi\chi\) production with the dark Higgs boson decay \shh.}
\label{fig:diag}
\end{figure}

The specific implementation of this model uses a two-mediator dark Higgs DM scenario, containing the dark Higgs boson, identified through its decay into \(hh\), a new broken \(U(1)'\) symmetry yielding an additional massive spin-1 \(Z'\) mediator, and a Majorana fermion DM candidate.
The model parameters are the DM particle mass \mchi; the \(Z'\) mass \mZprime; the dark Higgs boson mass \ms; the \(Z'\) coupling with quarks \gq\ and with the DM candidate \gchi; and the mixing angle \(\theta\) between the SM and dark Higgs bosons, which should be small to enable reproduction of the relic density and consistency with measurements of the Higgs boson.
The dominant \(q\bar q \to Z' \to s\chi\chi\) signal diagram for the large \(Z'\) masses explored in this paper can be found in Figure~\ref{fig:diag}.

A signal region is defined requiring significant \ptmiss\ and at least three hadronic jets identified to contain \(b\)-hadrons (\(b\)-tagged jets).
The dominant SM background processes in these regions are top quark pair production with additional radiation and \(Z\) boson production with multiple additional jets, where the \(Z\) boson decays invisibly into neutrinos.
Production of \(W\) bosons and single top quarks contribute sub-dominantly.
Control regions (CRs) are defined to constrain the yields of these processes in a partly data-derived approach.
A neural network (NN) is trained to improve the sensitivity through greater discrimination of signal from background processes whilst preserving the shape of the invariant mass distribution of the reconstructed di-Higgs system which is fit in the analysis.

Searches in similar final states have been made for supersymmetric particle cascades in which a pair of new heavy particles decay producing multiple \(b\)-quark jets and \ptmiss, either from heavy gluino or squark decay cascades~\cite{SUSY-2018-30,CMS-SUS-19-010} or electroweakino decays via the emission of Higgs bosons~\cite{SUSY-2020-16,CMS-SUS-20-004}.
The symmetric topology and lack of connection between the four \(b\)-quarks differs significantly from the resonant \(hh\) topology targeted in this paper.
Searches for dark Higgs bosons have been made in the \(bb\) signature by the ATLAS Collaboration~\cite{EXOT-2020-26}, and in \(WW\) final states by the ATLAS~\cite{EXOT-2020-04,EXOT-2018-40} and CMS~\cite{CMS-EXO-21-012} Collaborations.
The \(bb\) signature is sensitive to \(30 < \ms < 150\) \GeV, with \(WW\) sensitive from \(\ms = 150\) \GeV and growing weaker with increasing \ms, motivating the exploration of the \(hh\) decay for larger \ms\ values.
There is complementary sensitivity from spin-1 mediator searches compatible with the \(Z'\) mediator present in the standard benchmark model used to facilitate comparisons between signatures and experiments. Care is needed in translating them into the Majorana DM dark Higgs model, but limits from dijet searches~\cite{EXOT-2018-62-fixed} are strong and place powerful constraints on \TeV-mass \(Z'\) mediators, assuming the assumptions underlying the simplified models are fully realised in a more complete model, which may not be the case.
Generally, the balance of the resonant and \(\ptmiss\)\ signatures depends strongly upon the choice of the couplings \gq\ and \gchi, and the model is not strongly constrained. Reprodution of the observed relic density tends to favour larger \gchi\ values~\cite{Duerr:2017uap,EXOT-2020-26}, emphasising the importance of the \(X+\ptmiss\) signatures.

The paper is organised as follows: Section~\ref{sec:detector} describes the ATLAS detector, whilst Section~\ref{sec:dataMC} details the collision data and simulation used in the analysis.
Section~\ref{sec:eventReco} documents the object reconstruction and event selections used.
The analysis strategy is outlined in Section~\ref{sec:strategy}, before Section~\ref{sec:backgrounds} explains how background processes are estimated.
The systematic uncertainties evaluated on data and simulations are detailed in Section~\ref{sec:systs}, and the statistical treatment of the data in Section~\ref{sec:stats}.
The analysis results are reported in Section~\ref{sec:results}; Section~\ref{sec:conclusion} provides a conclusion.

%
%


%
\section{ATLAS detector}
\label{sec:detector}

\newcommand{\AtlasCoordFootnote}{%
ATLAS uses a right-handed coordinate system with its origin at the nominal interaction point (IP)
in the centre of the detector and the \(z\)-axis along the beam pipe.
The \(x\)-axis points from the IP to the centre of the LHC ring,
and the \(y\)-axis points upwards.
Polar coordinates \((r,\phi)\) are used in the transverse plane,
\(\phi\) being the azimuthal angle around the \(z\)-axis.
The pseudorapidity is defined in terms of the polar angle \(\theta\) as \(\eta = -\ln \tan(\theta/2)\) and is equal to the rapidity
$ y = \frac{1}{2} \ln \left( \frac{E + p_z}{E - p_z} \right) $ in the relativistic limit.
Angular distance is measured in units of \(\Delta R \equiv \sqrt{(\Delta y)^{2} + (\Delta\phi)^{2}}\).}
The ATLAS experiment~\cite{PERF-2007-01} at the LHC is a multipurpose particle detector
with a forward--backward symmetric cylindrical geometry and a near \(4\pi\) coverage in
solid angle.\footnote{\AtlasCoordFootnote}
It consists of a tracking inner detector (ID) surrounded by a thin superconducting solenoid
providing a \qty{2}{\tesla} axial magnetic field, electromagnetic and hadronic calorimeters, and a muon spectrometer.
The inner tracking detector covers the pseudorapidity range \(|\eta| < 2.5\).
It consists of silicon pixel, silicon microstrip, and transition radiation tracking detectors.
Lead/liquid-argon (LAr) sampling calorimeters provide electromagnetic (EM) energy measurements
with high granularity within the region \(|\eta|< 3.2\).
A steel/scintillator-tile hadronic calorimeter covers the central pseudorapidity range (\(|\eta| < 1.7\)).
The endcap and forward regions are instrumented with LAr calorimeters
for EM and hadronic energy measurements up to \(|\eta| = 4.9\).
The muon spectrometer surrounds the calorimeters and is based on
three large superconducting air-core toroidal magnets with eight coils each.
The field integral of the toroids ranges between \num{2.0} and \qty{6.0}{\tesla\metre}
across most of the detector.
The muon spectrometer includes a system of precision tracking chambers up to \(|\eta| = 2.7\) and fast detectors for triggering up to \(|\eta| = 2.4\).
The luminosity is measured mainly by the LUCID--2~\cite{LUCID2} detector, which is located close to the beampipe.
A two-level trigger system is used to select events~\cite{TRIG-2016-01}.
The first-level trigger is implemented in hardware and uses a subset of the detector information
to accept events at a rate below \qty{100}{\kHz}.
This is followed by a software-based trigger that
reduces the accepted event rate to \qty{1.25}{\kHz} on average
depending on the data-taking conditions.
A software suite~\cite{SOFT-2022-02} is used in data simulation, in the reconstruction
and analysis of real and simulated data, in detector operations, and in the trigger and data acquisition
systems of the experiment.


%
\section{Data and Monte Carlo simulations}
\label{sec:dataMC}

%
The proton--proton collisions analysed were collected by the ATLAS detector between 2015 and 2018, at a centre-of-mass energy of \(\sqrt s=13~\TeV\) and a 25~ns inter-bunch spacing.
They correspond to an integrated luminosity of \(140~\ifb\), with an uncertainty of 0.83\%~\cite{DAPR-2021-01} obtained using the LUCID-2 detector~\cite{LUCID2} for the primary luminosity measurements, complemented by measurements using the inner detector and calorimeters, after requiring all detector systems to be recording good quality data~\cite{DAPR-2018-01}.

Unprescaled \ptmiss\ triggers were used~\cite{TRIG-2019-01} to select events for the analysis.
The \ptmiss\ thresholds ranged from 70~\GeV\ to 110~\GeV, for a \ptmiss\ definition in which muons are treated as invisible.
The offline \ptmiss\ selection, which does not treat muons as invisible, was kept more stringent than the trigger-level requirement to ensure that trigger efficiencies are constant.

Dark Higgs Monte Carlo (MC) simulated events were generated, as detailed in Table~\ref{tab:datamc:mc}, using consistent parameter choices with other analyses~\cite{EXOT-2020-04,EXOT-2018-40,CMS-EXO-21-012,EXOT-2020-26} and simplified dark matter models: the dark matter mass \mchi\ is fixed at \(\SI{200}{\GeV}\), its coupling constant is set as \(\gchi=1\), the coupling between the SM quarks and the \Zprime\ is fixed at \(\gq=0.25\)~\cite{Abdallah:2015ter,Abercrombie:2015wmb} and the mixing angle between the SM and dark Higgs bosons is set to \(\sin{\theta}=0.01\).
Samples of MC simulated events are generated for 106 mass points covering a range of \mZprime\ and \ms\ values.
To consider the region where the \Zprime\ is heavy enough to decay into two dark matter particles and is light enough to have a sufficient production cross-section for sensitivity, \(\SI{420}{\GeV}<\mZprime<\SI{3300}{\GeV}\) is chosen.
The range \(\SI{251}{\GeV}<\ms<\SI{435}{\GeV}\) is chosen to probe the region where \(s\) is heavy enough to decay into two SM Higgs bosons and light enough that the decay into two dark matter particles in this benchmark is not overwhelming.
Larger DM masses will have generally smaller \Zprime\ and \(s\) production cross-sections, with \(s\rightarrow\chi\chi\) decays becoming dominant at larger \ms\ and a different predicted relic density.
For all generated signal events, the detector response is simulated~\cite{SOFT-2010-01} using a fast parametrised simulation of the calorimeters~\cite{SIMU-2018-04}, and the full \GEANT~\cite{Agostinelli:2002hh} simulation for the other sub-detectors.

The dominant SM background processes are \ttbar production with additional radiation, and \Znnjets.
Subdominant background processes are single top (\(Wt\)) production and \Wlnjets.
Smaller background processes include other multi-top processes hereinafter grouped as \ttX\ (\(ttW, ttZ, ttH\), four-top production), and diboson (\(WW, WZ, ZZ\)) production.
Triboson production and \(VH\) production were also considered and found to be negligible.
All SM processes are modelled using MC simulations, processed
through a full simulation of the ATLAS detector~\cite{SOFT-2010-01} based on \GEANT~\cite{Agostinelli:2002hh}.
All samples used (signal and SM) are listed in Table~\ref{tab:datamc:mc} along with the relevant parton distribution function (PDF) sets used for the matrix element (ME) and parton shower (PS), the configuration of underlying-event and hadronisation parameters (tune), and the cross-section order in \(\alpha_\mathrm{s}\) (and \(\alpha_\mathrm{EW}\) where corrections are available) used to normalise the event yields for these samples.
Further information about the ATLAS simulations of \ttbar, single top, multi-boson and vector-boson plus jets processes can be found in Refs.~\cite{ATL-PHYS-PUB-2018-009,ATL-PHYS-PUB-2016-004,ATL-PHYS-PUB-2017-005,ATL-PHYS-PUB-2017-006,PMGR-2021-01}. %
The decays of bottom and charm hadrons were simulated using the \EVTGEN[1.6.0] program~\cite{Lange:2001uf} for the samples not generated with \SHERPA.
The effect of pile-up in the same and neighbouring bunch crossings was modelled by overlaying the simulated hard-scattering event with inelastic proton--proton events generated with \PYTHIA{8.186}~\cite{Sjostrand:2007gs} using the \NNPDF{2.3LO} leading-order PDF set~\cite{Ball:2012cx} and the A3 set of tuned parameters~\cite{ATL-PHYS-PUB-2016-017}.
The MC samples were reweighted so that the distribution of the average number of interactions per bunch crossing reproduces the observed distribution in the data.

The \ttbar\ processes are categorised based on the true flavour content of the jets in the events.
An event is labelled as \ttbarb\ if three or more jets are associated with \(b\)-hadrons.
Events with two \(b\)-jets and at least one \(c\)-hadron are labelled as \ttbarc\, while all remaining events are categorised as \ttbarl.

\begin{table}[t]
\tiny
\centering
\caption{\label{tab:datamc:mc}Simulated event samples with the corresponding matrix element and parton shower generators, cross-section order in \(\alpha_\mathrm{s}\) (and \(\alpha_\mathrm{EW}\) where corrections are available) used to normalise the event yield, underlying-event set of tuned parameters and the generator PDF sets used. The abbreviation (N)NLO stands for (next-to-)next-to-leading-order, and NNLL stands for next-to-next-to-leading-logarithm.
}
\begin{tabular}{|p{0.1\linewidth}|p{0.17\linewidth}p{0.09\linewidth}p{0.13\linewidth}p{0.11\linewidth}p{0.11\linewidth}p{0.11\linewidth}|}
\hline
Physics process					& Generator (ME) 															& Parton shower (PS)										&  Order in \(\alpha_\mathrm{s}\)(\(+\alpha_\mathrm{EW}\))																											& Tune														& PDF (ME)										& PDF (PS)\\
\hline
Dark Higgs signals		& \MGNLO[2.9.9]~\cite{Alwall:2014hca}										& \PYTHIA[8.307]~\cite{Bierlich:2022pfr}			& LO																														& A14~\cite{ATL-PHYS-PUB-2014-021}							& \NNPDF[2.3LO]~\cite{Ball:2012cx}				& \NNPDF[2.3LO]\\
\Znnjets\					& \SHERPA[2.2.1]\cite{Bothmann:2019yzt,Gleisberg:2008fv}					& \SHERPA[2.2.1]									& NNLO~\cite{Gavin:2010az}																									& \SHERPA\ default~\cite{ATL-PHYS-PUB-2016-003}				& \NNPDF[3.0NNLO]~\cite{Ball:2014uwa}			& \NNPDF[3.0NNLO]\\
\Wlnjets\ 			& \SHERPA[2.2.1]															& \SHERPA[2.2.1]									& NNLO																														& \SHERPA\ default											& \NNPDF[3.0NNLO]								& \NNPDF[3.0NNLO]\\
Diboson					& \SHERPA[2.2.1]															& \SHERPA[2.2.1]									& NLO																														& \SHERPA\ default											& \NNPDF[3.0NNLO]								& \NNPDF[3.0NNLO]\\
Triboson				& \SHERPA[2.2.14]															& \SHERPA[2.2.14]									& NLO																														& \SHERPA\ default											& \NNPDF[3.0NNLO]								& \NNPDF[3.0NNLO]\\
\ttbar\		& \POWHEGBOX[v2]\cite{Re:2010bp,Nason:2004rx,Frixione:2007vw,Alioli:2010xd}	& \PYTHIA[8.230]~\cite{Sjostrand:2014zea}							& NNLO+NNLL~\cite{Czakon:2011xx}																							& A14														& \NNPDF[3.0NLO]~\cite{Ball:2014uwa}			& \NNPDF[2.3LO]\\
Single top	& \POWHEGBOX[v2]															& \PYTHIA[8.230]									& NLO~\cite{Aliev:2010zk,Kant:2014oha}																						& A14														& \NNPDF[3.0NLO]								& \NNPDF[2.3LO]\\
\(ttW/ttZ\)				& \MGNLO[2.3.3]											  					& \PYTHIA[8.210]									& NLO																														& A14														& \NNPDF[3.0NLO]								& \NNPDF[2.3LO]\\
Four-top					& \MGNLO[2.2.2]											  					& \PYTHIA[8.186]~\cite{Sjostrand:2007gs}									& NLO																														& A14														& \NNPDF[2.3LO]									& \NNPDF[2.3LO]\\
\(ttH\)					& \POWHEGBOX[v2]											  				& \PYTHIA[8.230]									& NLO																														& A14														& \NNPDF[3.0NLO]								& \NNPDF[2.3LO]\\
\(VH\)					& \PYTHIA[8.186]										  					& \PYTHIA[8.186]									& NNLO+NLO~\cite{Ciccolini:2003jy,Brein:2003wg,Brein:2011vx,Altenkamp:2012sx,Denner:2014cla,Brein:2012ne,Harlander:2014wda}	& A14														& \PDFforLHC[15NLO]~\cite{Butterworth:2015oua}	& \PDFforLHC[15NLO]\\
\hline
\end{tabular}
\end{table}


%
\section{Event reconstruction}
\label{sec:eventReco}

%
For the particle candidate objects used in this analysis, two distinct categories are defined.
These differ in the stringency of their reconstruction requirements, and have different uses in the analysis.
\baseline\ objects are used to construct the events' \ptmiss.
After this, an overlap removal procedure is applied to deal with potential ambiguities between particle candidates.
Finally \signal\ objects are defined with additional reconstruction requirements, which are used for the analysis region selections.

All events are required to have a primary vertex.
Primary vertices are reconstructed from at least two charged-particle tracks with transverse momentum \(\pt>\SI{0.5}{\GeV}\), measured in the ID~\cite{ATL-PHYS-PUB-2015-026}.
The hard-scatter vertex is defined as the primary vertex with the largest sum of squared track momenta.

Hadronic jets are reconstructed from particle flow objects~\cite{PERF-2015-09}, using the \antikt\ algorithm~\cite{Fastjet,Cacciari:2008gp} with a radius parameter of \(R=0.4\).
These are calibrated using a multi-step procedure as outlined in Ref.~\cite{JETM-2018-05}.
\baseline\ jets are required to have \(\pt>\SI{20}{\GeV}\) and \(|\eta|<2.8\) to be considered in the analysis.
To suppress jets originating from pile-up, a Jet Vertex Tagger (\texttt{JVT}) is used; this is a likelihood-based algorithm described in Ref.~\cite{PERF-2014-03}.
\baseline\ jets with \(20 < \pT < \SI{60}{\GeV}\) and \(|\eta|<2.4\) are required to satisfy a selection on the \texttt{JVT} with a \(0.90\)--\(0.96\) probability of correctly identifying hard-scatter jets depending on the \pT.
Outside of this \pT\ range, no \texttt{JVT} requirement is applied.
\signal\ jets are formed from the \baseline\ jets which pass the overlap removal procedure, and are additionally required to satisfy an increased \pT\ requirement of \(\SI{25}{\GeV}\).

For both \baseline\ and \signal\ jets, those likely to originate from a \(b\)-quark are identified using a deep neural network-based algorithm (\texttt{DL1r})~\cite{FTAG-2019-07}.
This is applied to jets with \(\pt>\SI{20}{\GeV}\) and \(|\eta|<2.5\), with a \(77\%\) probability to identify jets containing \(b\)-hadrons, as measured in MC simulations of \ttbar\ processes.
The algorithm has a probability of \(0.5\%\) (\(18\%\)) of misidentifying light-flavour (charm) jets, as determined in a sample of simulated \ttbar\ events.
The \(b\)-tagging efficiencies in simulated events are corrected to match those measured in data~\cite{FTAG-2018-01,FTAG-2020-08,FTAG-2019-02}.

\baseline\ electrons (muons) are required to have \(|\eta|<2.47\) (\(|\eta|<2.7\)) and \(\pt>\SI{8}{\GeV}\), and to satisfy the \loose\ likelihood-based identification working point defined in Ref.~\cite{EGAM-2018-01} (\medium\ identification working point as defined in Ref.~\cite{MUON-2018-03}).
\signal\ electron and muon candidates are defined to place a stringent veto on events with any prompt leptons, to reduce SM backgrounds, and \signal\ muons are used in the estimate of the \(Z\)-boson background.
These are required to fulfil the following additional requirements and the overlap removal procedure must be passed.
The significance of the transverse impact parameter \(d_0\) must obey \(d_0/\sigma(d_0)\) less than 5 (3), for electrons (muons).
\signal\ electrons are required to satisfy the \loose\ isolation working point of Ref.~\cite{EGAM-2018-01} to reduce contamination from semileptonic hadron decays and jets misidentified as electrons.
\signal\ muons are required to satisfy the \texttt{Tight} isolation working point of Ref.~\cite{MUON-2018-03} to reduce contamination from semileptonic heavy flavour and hadron decays.

The overlap removal procedure is applied to \baseline\ electrons (\(e\)), muons (\(\mu\)) and jets (\(j\)).
Individual overlap removals are done in the following sequence of steps, in which objects are considered at a given step only if they were not removed in a previous step.
Firstly, if a muon shares an ID track with an electron, the electron is removed.
Secondly, if \(\Delta R(j, e) < 0.2\) the jet is removed.
For any surviving jets, if \(\Delta R(j, e) < \)~min(0.4, 0.04 + \SI{10}{\GeV}/\(p^{e}_\text{T}\)), the electron is removed.
Penultimately, if \(\Delta R(j, \mu) < 0.2\) or the muon ID track is ghost associated\footnote{Ghost association consists of repeating the \antikt\ jet clustering process with the addition of `ghost' versions of the muon's track with the
same direction but infinitesimal \pt. A muon is ghost-associated if it is contained within the re-clustered jet.} with the jet, then the jet is removed if the jet has less than three associated tracks with \(\pt >\) 0.5~\GeV.
Finally, for any surviving jets, if \(\Delta R(j, \mu) < \) min(0.4, 0.04 + 10 \GeV/$p^{\mu}_{\text{T}}$), the muon is removed.

The missing transverse momentum (\(\ptmissvec\)) of the event is defined as the negative vector sum of the \(\ptvec\) of all selected and calibrated \baseline\ physics objects in the event before the overlap removal procedure, with an extra term added to account for soft energy in the event which is not associated with any of the selected objects.
This soft term is calculated from inner detector tracks matched to the primary vertex.
The magnitude of the missing transverse momentum vector is denoted \(\ptmiss\).
The \loose\ \ptmiss\ working point is used, as defined in Ref.~\cite{JETM-2020-03}.
The object-based \ptmiss\ significance (\metsig), defined in Ref.~\cite{JETM-2020-03}, is also used in the analysis.
This is an estimate of how genuine the \ptmiss\ in the event is, originating from neutrinos or potential dark matter candidates, in contrast to `fake' \ptmiss\ originating from the finite detector acceptance and resolution, pile-up and object mismeasurement. It is calculated by dividing the \ptmiss\ by a measure of its resolution, built from the resolutions of the objects from which the \ptmiss\ is constructed.


%
\section{Analysis strategy}
\label{sec:strategy}

\subsection{Event selection}
\label{sec:strategy:selection}

The signal is characterised by large \ptmiss\ from the DM particle production and substantial hadronic activity from \shh\ decays resulting in four \(b\)-quark jets with an invariant mass consistent with \ms.
Events considered for the search must fulfil a set of `preselection' requirements summarised in Table~\ref{tab:presel}.
To target the signal's final state, events are required to contain no \signal\ leptons (\nleptons) and between four and seven \signal\ jets (\njets), at least three of which must be \(b\)-tagged. The loose upper requirement on \njets\ reduces backgrounds with many additional jets, including top quark processes.
Requiring the events' \ptmiss\ to be greater than \(\SI{200}{\GeV}\) ensures the selected trigger is fully efficient over the complete data-taking period, and effectively targets the high \ptmiss\ signature for the range of DM masses considered.
To reduce background contributions from QCD multijet processes, \dPhiMin\ --- the minimum azimuth between any of the event's four highest \pt\ jets and the \ptmissvec --- is required to be greater than 0.4.
The variable \mTbmin, defined as:
\begin{equation*}
\mTbmin = \min_{1\leq i\leq 3}\sqrt{(\ptbX{i}+\ptmiss)^2-(\pxbX{i}+\pxmiss)^2-(\pybX{i}+\pymiss)^2},
\end{equation*}
where \(i\) indexes the three highest-\pT\ \(b\)-tagged jets in the event, has an approximate upper bound at the mass of a particle with decay products including a \(b\)-tagged jet and \ptmiss, such as a semileptonic top quark decay in which the lepton is not reconstructed.
It is thus generally smaller for SM background processes than the signal.
The optimal event requirement was found to be \mTbmin\ larger than \(\SI{80}{\GeV}\).
To reduce backgrounds with fake \ptmiss\ such as QCD multijet processes, events must have a \metsig\ greater than 7.
Finally, the transverse mass of the di-Higgs system and \ptmiss must exceed \(\SI{300}{\GeV}\) to target events consistent with a heavy \(Z'\) decay.

\begin{table}[t]
\centering
\caption{Definition of the event preselection requirements, used to select the final states of interest.}
\begin{tabular}{|l|c|}
\hline
Variable		    & Requirement \\
\hline
\nleptons\  	    & 0 \\
\njets\			    & \([4,7]\) \\
\nbjets\		    & \(\geq 3\) \\
\ptmiss\ \([\GeV]\)	& \(>200\) \\
\dPhiMin		    & \(>0.4\) \\
\mTbmin\ \([\GeV]\) & \(>80\) \\
\metsig				& \(>7\) \\
\mTfourb\ \([\GeV]\)	& \(>300\) \\
\hline
\end{tabular}
\label{tab:presel}
\end{table}

To discriminate further between the signals and SM backgrounds, an NN classifier is used, as detailed in Section~\ref{sec:strategy:ML}, which produces a probability score for events ranging from \(0\) (highly background compatible) to \(1\) (highly signal compatible).
For all the search regions in the analysis, the events' NN output score is required to be greater than 0.95 to maximise sensitivity to the dark Higgs signals whilst keeping sufficient event yields to maintain the clear resonant peak signature expected in the signal in the di-Higgs invariant mass (\mhh).
For the dark Higgs model this corresponds to the reconstructed \ms, and has a resonance peak in its distribution.

Two sets of search regions are defined in Table~\ref{tab:SRs}, which both exploit \mhh\ to distinguish the signals from the SM backgrounds.
Firstly, a binned signal region (\SR), optimised for sensitivity to the dark Higgs signal resonances, is used to test the signal hypotheses in multi-bin profile likelihood fits over \mhh.
The \mhh\ distribution is binned with the set of lower edges: \([100, 200, 250, 300, 350, 400, 450, 500, 700] ~\GeV\), with the final bin ending at \(\SI{900}{\GeV}\).
The binning is chosen to reflect the mass resolution of the dark Higgs signal and there are no yields in data or MC simulation above \(\SI{900}{\GeV}\).
The overall acceptance times efficiency for the dark Higgs signals to enter the \SR\ varies from \(1.5\%\) to \(6.9\%\) across the parameter space considered.
Secondly, a set of single bin regions (\DRa-\DRg) is used to perform individual discovery fits testing the SM-only hypothesis and set limits on the possible number of BSM events with this detector signature.
These are each defined in a given range of \mhh.
These overlapping regions are chosen to cover a range of possible BSM particle masses, contain sufficient expected event yields, and reflect the expected mass resolution.
Whilst the 4\(b\) final state is the most sensitive at preselection level, the larger event samples in the 3\(b\) regions benefit most from the NN discrimination, such that both contribute and a categorisation in \(b\)-tagged jet multiplicity does not enhance the sensitivity.

\begin{table}[t]
\centering
\caption{Search region event requirements supplementing the preselection defined in Table~\ref{tab:presel}.}
\begin{tabular}{|l|c|c|}
\hline
Region		    & \multicolumn{2}{c|}{Variable requirement} \\
& NN score						& \mhh\ \([\GeV]\) \\
\hline
\SR\			& \(\geq 0.95\)					& \([100,900]\)\\
\hline
\DRa\			& \multirow{7}{*}{\(\geq0.95\)}	& \([100,250)\) \\
\DRb\			& 								& \([200,300)\) \\
\DRc\			& 								& \([250,350)\) \\
\DRd\			& 								& \([300,400)\) \\
\DRe\			& 								& \([350,450)\) \\
\DRf\			& 								& \([400,900)\) \\
\DRg\			& 								& \([500,900)\) \\
\hline
\end{tabular}
\label{tab:SRs}
\end{table}

\subsection{Higgs boson reconstruction}
\label{sec:strategy:Higgs}

The fully visible final state of the dark Higgs boson decay allows its mass to be reconstructed, providing an effective discriminant between the signal process and SM backgrounds.
This relies first on selecting the four jets from the two SM Higgs boson decays from which the invariant mass (\(m_{4j}=\mhh\)) can be calculated.
If there are at least four \(b\)-tagged jets in the event, the four \(b\)-tagged jets with the largest transverse momenta are selected.
If there are three \(b\)-tagged jets, and the mass of one of these is greater than \(\SI{100}{\GeV}\), then it is assumed to be a `merged' jet formed from two Higgs boson decay products being collimated, and used in addition to the other two \(b\)-tagged jets.
If the mass threshold is not met by any of the \(b\)-tagged jets, then the fourth jet is chosen from the remaining untagged jets.
This jet is chosen to minimise the mass of the heaviest possible Higgs boson candidate constructed from any two of the set containing the three \(b\)-tagged jets (\(\alpha, \beta, \gamma)\) and the considered light jets (\(j\)):
\begin{equation*}
\min_{j}
\Biggl[
\max_{\alpha, \beta, \gamma}
\Bigl(
m_{\alpha \beta}, m_{\gamma j}
\Bigr)
\Biggr] .
\end{equation*}

The kinematic properties of the SM Higgs boson candidates are also useful for discriminating between the signal and backgrounds and are used as input features to the NN\@.
To reconstruct these, the four jets must be paired to form two Higgs bosons.
When there is a merged jet present, it is chosen as one Higgs boson candidate, whilst the other two jets form the second Higgs boson candidate.
Otherwise, the four jets (\(j=1,2,3,4\)) are paired to minimise the greatest angular separation between the Higgs boson candidates decay products (jets \(i,k,l,m\)).
This is expressed as:
\begin{equation*}
\text{min}(\Delta R_{\text{max}}(b,b)) = \text{min}_{i,k,l,m \in j}(\text{max}(\Delta R(i,k), \Delta R(l,m))).
\end{equation*}
The two Higgs boson candidates are then numbered in descending \pT.

\subsection{Neural network classifier}
\label{sec:strategy:ML}

The supervised NN classifier is trained on events satisfying the preselection requirements defined in Table~\ref{tab:presel}.
For the signal training set, simulated MC events for all signal models are combined (using the same number of events per model, randomly sampled such that the $4.3\times10^{5}$ signal training set size is equal to that of the background training set), to maintain sensitivity across the model parameter space of interest.
For the background training set, all MC simulated events of the SM background processes were used as they provided a good match to the expected weighted background composition satisfying the analysis preselection.
To avoid overtraining and benefit from the full set of available events, orthogonal training and testing/validation datasets are used, through a \(k\)-fold splitting procedure with \(k=5\)~\cite{kfold}.
The datasets are split randomly into five subsets and five NNs are trained, such that the network trained on four fifths of the events is applied to the other fifth of the events in the analysis, and the full statistical power of the simulated MC datasets can be conserved.

The kinematic input features used in the NN are listed and defined in Table~\ref{tab:NNinputs}.
These were selected from a wider set as the kinematic variables that provided the best signal-background discrimination and improved the NN performance, either alone or in combination with other variables.
The NN can exploit correlations between variables to enhance discrimination power, particularly when the correlations between the variables are different for signal and background events.
The impact of each variable was also quantified using Shapley values~\cite{Shapley} (a common measure to assess how impactful an input feature is on the classifier output).
Variables which had high Shapley values, and led to a reduced NN performance if they were removed from training, were retained.
Finally, more complex compound kinematic variables were considered and evaluated, again using their Shapley values.
These variables were generated using the Python \texttt{featuretools}~\cite{kanter2015deep} package, which combines pairs of variables from the existing list using basic arithmetic operations.
The compound variables with high Shapley values, and which improved NN performance if they were added, were included in the training.
Whilst these variables are correlated with the existing input features, they can help the NN to converge more quickly, focus more on more important properties, and learn finer details in the data sample.

\begin{table}[tp]
\centering
\caption{Neural network input features and their definitions.}
\begin{tabular}{|l|p{12cm}|}
\hline
Variable						& Definition \\
\hline
\ptmiss							& Missing transverse momentum. \\
\metsig							& Object-based \ptmiss\ significance. \\
\mTbmin							& Minimum transverse mass of \ptmiss\ and any of the $b$-jets in the event. \\
\mTfourb						& Transverse mass of the di-Higgs system and \ptmiss. \\
\mhX{1}							& Mass of the higher-\pT\ Higgs boson candidate. \\
\mhX{2}							& Mass of the lower-\pT\ Higgs boson candidate. \\
\HT\							& Scalar sum of \pt\ of \signal\ jets. \\
\ptX{4b}						& \pT\ of the reconstructed di-Higgs system. \\
\ptbX{1}						& Transverse momentum of the highest \pT\ $b-$jet. \\
\ptbX{2}						& Transverse momentum of the second-highest \pT\ $b-$jet. \\
\ptX{j4}						& Transverse momentum of the fourth-highest \pT\ jet. \\
\dRHbb{1}						& $\Delta R$ between the two jets associated with the higher-\pT\ Higgs boson candidate. \\
\dRHbb{2}						& $\Delta R$ between the two jets associated with the lower-\pT\ Higgs boson candidate. \\
\dRminbb						& \(\min~\Delta R(b_i,b_j)\) for all $b$-jet pairs \(i\) and \(j\) in the event. \\
\dEtaMaxbb						& \(\max~\Delta \eta(b_i,b_j)\) for all $b$-jet pairs \(i\) and \(j\) in the event. \\
\njets							& Number of \signal\ jets in the event. \\
\nbjets							& Number of $b$-jets in the event. \\
$\ptmiss / \sqrt{\HT}$			& Event-based (rather than object-based) estimate of \metsig. \\
$\ptmiss / \ptX{4b}$			& The ratio of \ptmiss\ and \ptX{4b}. \\
$\njets / \nbjets$				& The ratio of the \signal\ jet and \(b\)-jet multiplicities. \\
$\njets-\nbjets$				& The number of \signal\ jets in the event not \(b\)-tagged. \\
$\ptbX{1}/\njets$				& The ratio of the \pT\ of the highest \pT\ jet and the event's \signal\ jet multiplicity. \\
$\metsig/\nbjets$				& The ratio of \ptmiss\ significance and the number of \(b\)-jets in the event. \\
$\metsig\times\ptX{4b}$			& The product of \ptmiss\ significance and the \pT\ of the di-Higgs system. \\
\(\dRminOverMET\)				& The ratio of \dRminbb and \ptmiss. \\
$\dEtaRMSbb\times\dEtaMaxbb$	& The product of \(\text{RMS}~\Delta \eta(b_i,b_j)\) and \(\max~\Delta \eta(b_i,b_j)\) for all \(b\)-jet pairs \(i\) and \(j\) in the event. \\
$\dRminbb\times\ptbX{2}$		& The product of \dRminbb\ and the second-highest \(b\)-tagged jet \pT. \\
\texttt{xwt}					& \(\min~ \sqrt{\biggl( \frac{m_W^i - 80.4 ~\text{\GeV}}{0.1 m_W^i} \biggl)^2 + \biggl( \frac{m_t^i - 172.5 ~\text{\GeV}}{0.1 m_t^i} \biggl)^2 }\), for \(i\) in the top quark and \(W\)-boson candidates in the event. Top quark candidates are defined through a \(W\)-boson candidate and \(b\)-tagged jet such that each jet from a Higgs boson decay represents the \(b\)-jet from the top decay. \(W\)-boson candidates are defined from pairs of the event's remaining jets, where at most one is associated with a Higgs boson (as hadronic \(W\) decays are likely to produce at least one light jet). The 0.1 approximates the fractional mass resolution for the reconstructed particle candidate hypotheses. \\
\hline
\end{tabular}
\label{tab:NNinputs}
\end{table}

A feed-forward neural network architecture is used.
The NN has 8 fully connected hidden layers, with 1024 hidden nodes in the first layer and the number of hidden nodes halving repeatedly in each subsequent layer.
Training is performed in batches, with a batch size of 4096.
Rectified linear unit (ReLU) activation~\cite{FukushimaRelu,HintonRelu}, batch normalisation~\cite{batchnorm}, an Adam~\cite{kingma2017adam} optimiser with weight decay, and dropout regularisation~\cite{dropout} (with a dropout rate of 0.2) are used between layers.
An initial learning rate of \(5\times 10^{-6}\) is used, with a learning rate scheduler that halves the learning rate every five epochs.
The network is trained for 25 epochs; that with the lowest validation loss is used in the analysis.
As the signal regions in the analysis rely on the \mhh\ distribution alongside high NN score values, it is important that the NN does not sculpt or bias the background \mhh\ distribution to a more signal compatible shape.
To prevent this, the loss function used contains not only a standard binary cross-entropy term \(L_{\text{BCE}}\), but also a distance correlation term \(\mathcal{R}\) defined between vectors \(X\) and \(Y\) as:
\begin{equation*}
\mathcal{R}^2(X,Y) =
\begin{cases}
\frac{\mathcal{V}^2(X, Y)}{\sqrt{\mathcal{V}(X, X)\mathcal{V}(Y, Y)}}, & \text{if } \mathcal{V}^2(X)\mathcal{V}^2(Y)>0,
\\
0, & \text{otherwise},
\end{cases}
\end{equation*}
where
\begin{equation*}
\mathcal{V}^2(X,Y) = \langle |X-X'||Y-Y'| \rangle
+ \langle |X-X'|\rangle\langle|Y-Y'| \rangle
- 2\langle |X-X'||Y-Y''| \rangle,
\end{equation*}
following Refs.~\cite{distanceCorrelation} and~\cite{Kasieczka:2020yyl}.
These are combined thus~\cite{Kasieczka:2020yyl}:
\begin{equation*}
L_{\text{total}}(\hat{Y}, Y) = \alpha L_{\text{BCE}}(\hat{Y}, Y) + (1-\alpha) \mathcal{R}(\hat{Y}, \mhh),
\end{equation*}
where \(Y\) is the vector of true signal or background labels for the events, and \(\hat{Y}\) the predictions.
The optimal value of \(\alpha\) was found to be 0.7, providing successful mass decorrelation and good training and classification performance.

Some example distributions of discriminating kinematic variables important to the NN are shown in Figure~\ref{fig:kinematics}.
Figures~\ref{fig:kinematics:1} and \ref{fig:kinematics:2} show the \ptmiss\ and \metsig\ distributions.
The illustrated signals demonstrate the large \ptmiss\ and \metsig\ generated by the dark matter production, in comparison with the rapidly dropping distributions for SM processes, particularly \ttbar\ processes.
The \dEtaMaxbb\ distribution in Figure~\ref{fig:kinematics:3} demonstrates the discrimination resulting from the closer proximity of the \(b\)-tagged jets in the signal process compared with those in the background processes.
The \mTbmin distribution (Figure~\ref{fig:kinematics:4}) connects the \ptmiss\ and \(b\)-tagged jets: for semileptonic top quark decays it is constrained to lie within \mtop, whereas at high values processes with additional or independent sources of \ptmiss\ are increasingly dominant.

\begin{figure}[t]
\centering
\subfloat[]{\label{fig:kinematics:1}\includegraphics[width=0.49\textwidth]{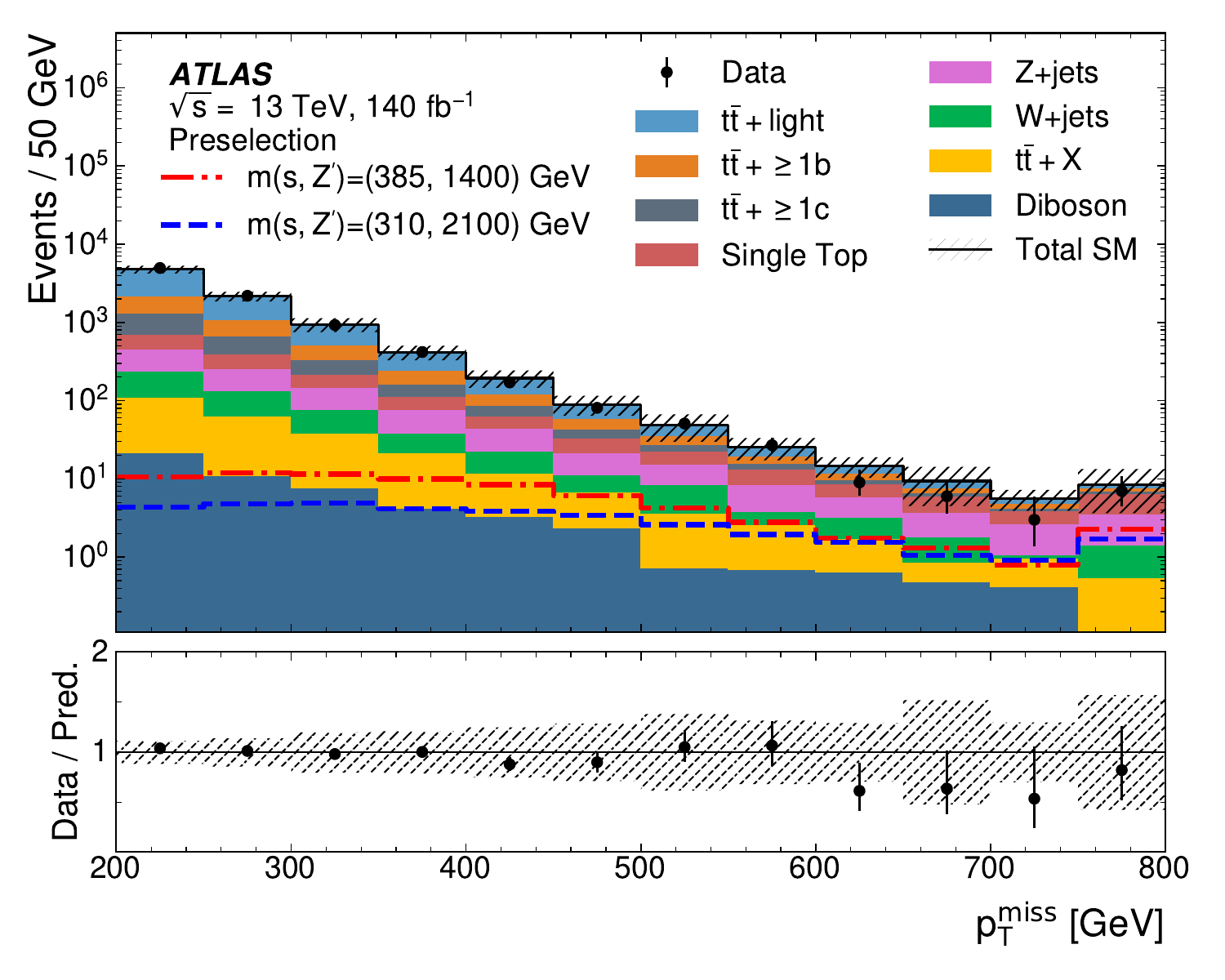}}
\subfloat[]{\label{fig:kinematics:2}\includegraphics[width=0.49\textwidth]{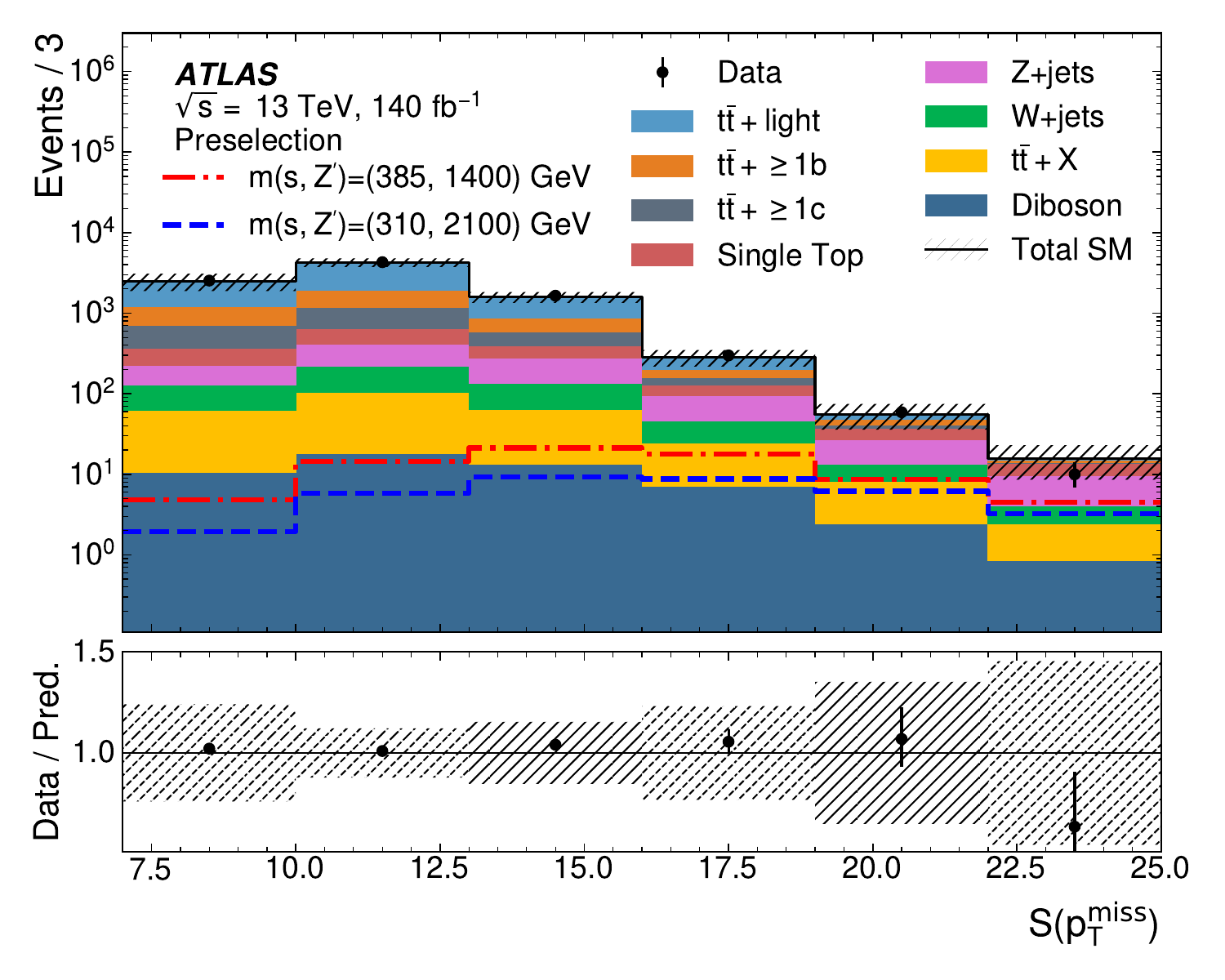}}\\
\subfloat[]{\label{fig:kinematics:3}\includegraphics[width=0.49\textwidth]{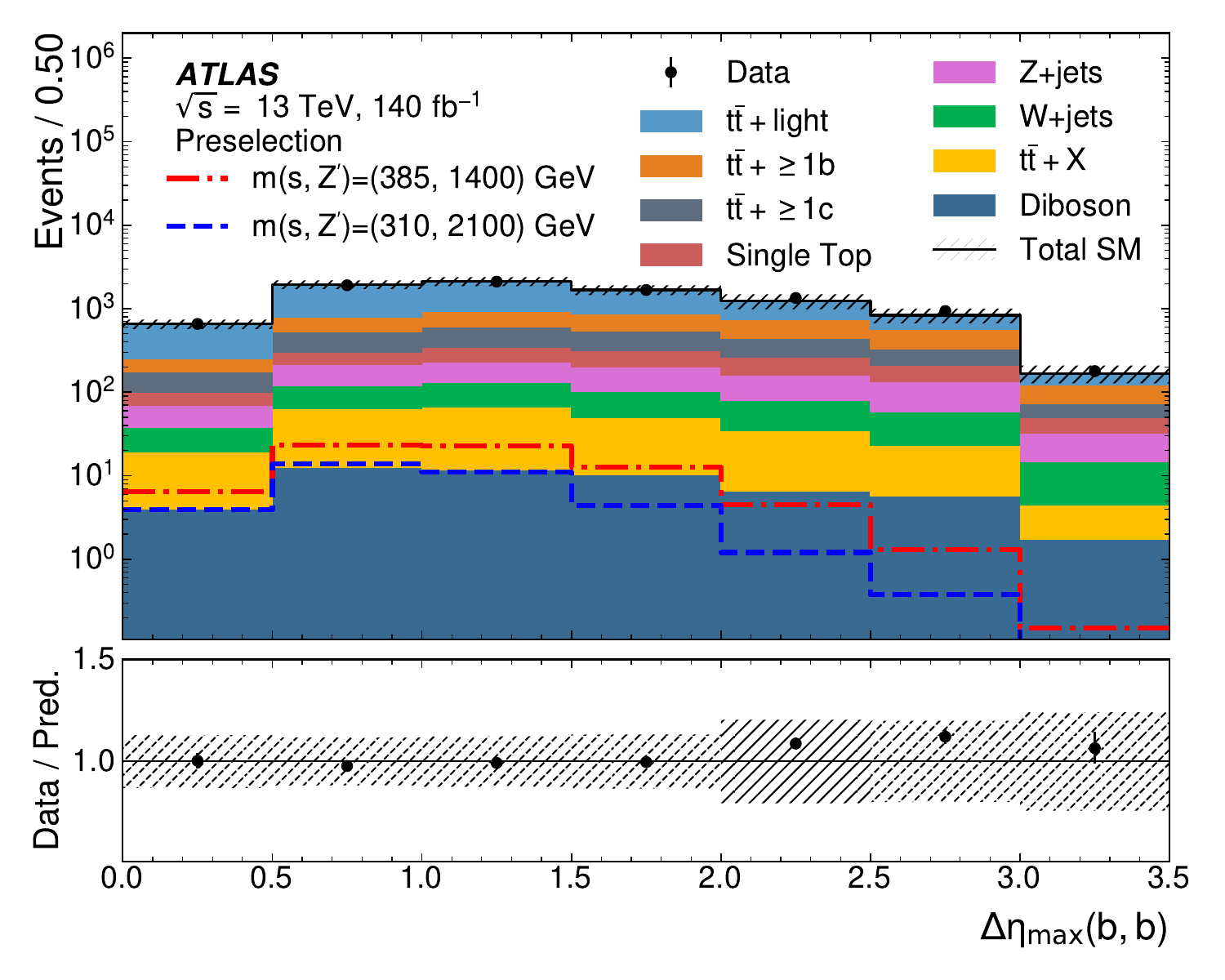}}
\subfloat[]{\label{fig:kinematics:4}\includegraphics[width=0.49\textwidth]{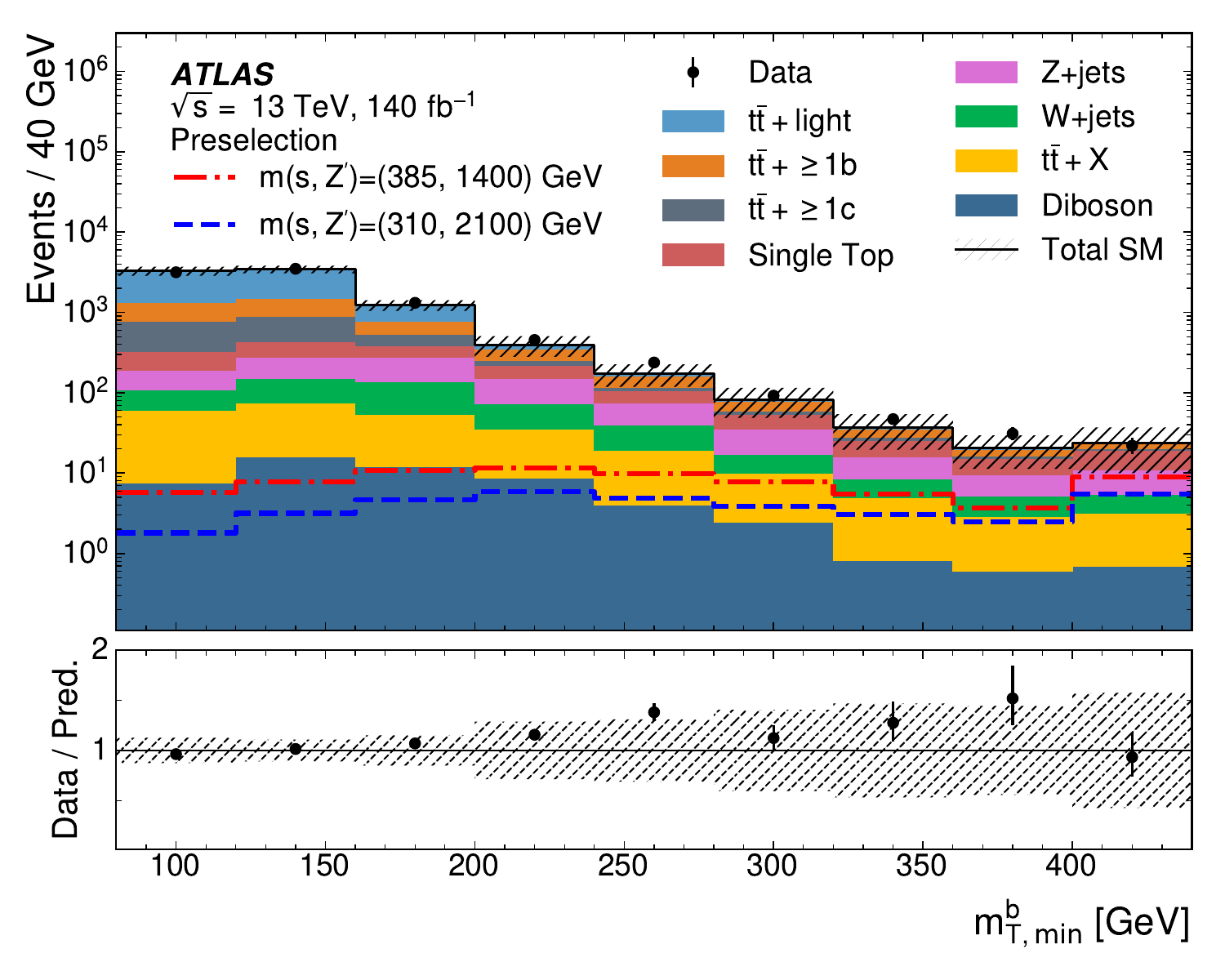}}\\
\caption{In the top panel of each plot, distributions of important kinematic variables in the NN training are given: (a) \ptmiss, (b) \ptmiss\ significance, (c) \dEtaMaxbb\ and (d) \mTbmin.
The preselection requirements (defined in Table~\ref{tab:presel}) are applied and the background, benchmark signals and data distributions are shown, before any fit is performed. The uncertainty band on the background includes all systematic and MC statistical uncertainties. In the bottom panels the ratio of data to background is shown. The final bin in each distribution includes overflow.}
\label{fig:kinematics}
\end{figure}


%
\section{Background estimation}
\label{sec:backgrounds}

%
After the preselection and search region NN score requirements, the dominant SM background processes remaining include \ttbar\ with additional radiation and \Zjets\ production with multiple additional jets.
The \ttbar\ backgrounds are divided into distinct processes depending on the flavour of the jets produced in association with the \ttbar\ process: \ttbarl, \ttbarb\ (which are both large) and \ttbarc\ (which is smaller).
The \Zjets, \ttbarl\ and \ttbarb\ backgrounds are estimated by using MC simulated events and dedicated control regions are defined to improve their modelling and provide normalisation factors derived from data.
These control regions are designed to be kinematically similar to the search regions, reasonably pure in the relevant background processes and low in signal contamination.\footnote{A signal contamination below 10\% for dark Higgs models not excluded by other searches is achieved in all control and validation regions.}
Sub-dominant background contributions come from \Wjets, single top, \ttX\ and diboson production processes.
These, and the \ttbarc\ background, are estimated with MC simulated events.
The QCD multijet background satisfying the preselection requirements is negligible.

The \Zjets\ process entering the search regions consists primarily of \Znn\ decays.
To obtain a high purity control region, a \Zmm\ selection is used.
This event sample is kinematically analogous to the signal process, contains high-\pT \(Z\) bosons orthogonal to the signal regions, and is treated identically by the \ptmiss\ triggers, which consider muons to be invisible.
So that the \Zmm\ events emulate the \Znn\ events in the search regions, their missing transverse momentum is calculated treating the muons as invisible, termed \metmu.
The control region definition starts by altering the standard preselection to require two \signal\ muons, and using \metmu\ instead of \ptmiss\ for the \(\SI{200}{\GeV}\) selection and \mTbmin calculation (\mTbminmu).
An event-based proxy of the muon-corrected missing transverse momentum significance is used (\(\metmu/\sqrt{\HT}\)), with the same lower bound of 7.
Events are required to have an unmodified \ptmiss\ below \(\SI{75}{\GeV}\) to reduce top quark backgrounds.
High purity in \Zjets\ is ensured by requiring that the muons are consistent with a \(Z\) boson decay, with an \mmumu\ invariant mass within \(\SI{15}{\GeV}\) of \mZ.
This CR has relatively limited yields but the predicted \Zjets\ background was found to reproduce the \SR\ kinematics and \mhh\ distribution well, avoiding the need to mirror the NN requirement by developing analogous NN input features based on \metmu.
The resulting \Zjets\ purity in \CRZ\ is \(\sim 80\%\), with a sub-leading contribution from \ttZ.
A summary of the \CRZ\ definition is given in Table~\ref{tab:CRZdef}.

\begin{table}[t]
\centering
\caption{Selection for \Zjets\ control region.}
\begin{tabular}{|l|c|}
\hline
Variable                   			& \CRZ\ Requirement \\
\hline
\nmu\                      			& \(=2\) \\
\njets\                     		& \([4,7]\) \\
\nbjets\                    		& \(\geq 3\) \\
\dPhiMin                    		& \(>0.4\) \\
\metmu\ \([\GeV]\)          		& $>200$ \\
\(\metmu/\sqrt{\HT}~~[\sqrt{\GeV}]\) & $>7$ \\
\mTbminmu \([\GeV]\)        		& $>80$ \\
\ptmiss\ \([\GeV]\)            		& $<75$ \\
\(|\mmumu - \mZ|~[\GeV]\)   		& $<15$ \\
\hline
\end{tabular}
\label{tab:CRZdef}
\end{table}

To estimate the \ttbarl\ and \ttbarb\ backgrounds, control regions (summarised in Table~\ref{tab:CRVRdef}) are defined to fulfil the preselection requirements and have a NN score in the 0.35 to 0.85 range to reduce signal contamination whilst remaining kinematically similar to the search regions.
The \ttbarb\ component is targeted by the \CRfourb\ region, requiring exactly four \(b\)-tagged jets, which increases its fractional yield to \(\sim45\%\).
To target the \ttbarl\ background, a control region requiring exactly three \(b\)-tagged jets is defined. To further increase its purity, an \(\SI{80}{\GeV}<\mTbmin<\SI{200}{\GeV}\) requirement is made, ensuring the \mTbmin\ value is compatible with the \(m_t\) endpoint. The resulting \CRthreeblow\ region, is \(\sim57\%\) pure in \ttbarl, with secondary contributions from other top quark background processes.
Its complement, with \mTbmin\ above \(\SI{200}{\GeV}\) and exactly three \(b\)-tagged jets forms the control region \CRthreebhigh.
This is used to check the %
modelling of background components contributing beyond \(\mTbmin = \mtop\), and has larger contributions from \Zjets\ (\(\sim 25\%\)), single top quark processes (\(\sim 20\%\)) and \ttbarb\ events (\(\sim 16\%\)).

To validate that the normalisation factors derived from the control regions lead to a good agreement between data and the background prediction in a signal-depleted region, a validation region (\VR) is defined.
This is made kinematically similar to the search regions using the adjacent 0.85 to 0.95 NN score region and the same preselection requirements, as shown in Table~\ref{tab:CRVRdef}.

\begin{table}[t]
\centering
\caption{Selections for all single top and \ttbar\ CRs, and the VR.}
\begin{tabular}{|l|ccc|c|}
\hline
Variable			& \multicolumn{4}{c|}{Region requirement} \\
& \CRthreeblow\ & \CRthreebhigh\    & \CRfourb\ & \VR\ \\
\hline
Preselection		& \multicolumn{3}{c|}{pass}                     & pass \\
NN score			& \multicolumn{3}{c|}{$[0.35,0.85)$}            & $[0.85,0.95)$ \\
\nbjets\			& $=3$         & $=3$             & $=4$     & \(\geq 3\) \\
\mTbmin\ $[\GeV]$	& $[80,200)$    & $\geq200$			& $\geq80$	& $\geq80$ \\
\hline
\end{tabular}
\label{tab:CRVRdef}
\end{table}


%
\section{Systematic uncertainties}
\label{sec:systs}

%
Experimental uncertainties in luminosity, \pileup\ reweighting, jet energy scale and resolution, \texttt{JVT} efficiency, \ptmiss\ soft term scale and resolution, and \(b\)-jet tagging efficiency are considered.
The \ptmiss\ soft term uncertainty is estimated following the procedure described in Ref.~\cite{JETM-2020-03}.
Uncertainties in the jet energy scale and resolution and \(b\)-jet tagging efficiency are estimated at a per-jet level from calibrations and comparisons between different event simulation models~\cite{PERF-2015-09,JETM-2018-05, FTAG-2018-01}.
The uncertainty in the \texttt{JVT} efficiency is estimated following the technique in Ref.~\cite{PERF-2014-03}.
Uncertainties related to lepton reconstruction and identification are found to be negligible.

Systematic uncertainties associated with the MC generator choice, parton shower, and the modelling of initial and final state radiation are considered for the \ttbar\ and single top backgrounds.
The generator (parton shower) uncertainty is evaluated by comparing the nominal \POWHEG+\PYTHIA[8] set-up against an alternative sample generated with \MGNLO+\PYTHIA[8] (\POWHEG+\HERWIG[7]).
The uncertainty due to initial-state radiation is estimated by simultaneously varying the \hdamp parameter and the \muR and \muF scales, and choosing the Var3c up/down variants of the A14 tune as described in Ref.~\cite{ATL-PHYS-PUB-2017-007}.
The final-state radiation uncertainty is evaluated by varying the renormalisation scale up or down by a factor of two.
The uncertainty associated with modelling the interference between \ttbar\ and single top processes is evaluated by comparing the nominal diagram removal~\cite{Frixione:2008yi} set-up to an alternative sample with diagram subtraction~\cite{Frixione:2008yi,ATL-PHYS-PUB-2016-020}.

Uncertainties related to missing higher-order corrections are assessed by variations of the choice of renormalisation and factorisation scales for the
\Wjets\ and \Zjets\ backgrounds, evaluated as the envelope of a set of seven variations~\cite{Bothmann:2016nao}.
Uncertainties in the resummation and matching were also assessed and found to be negligible.
A conservative \(50\%\) uncertainty is used for the \ttbarc\ and single top backgrounds to account for the uncertainty in their cross-sections in the signal phase space~\cite{SUSY-2020-16}, which has negligible impact on the results.
For the minor \ttX\ and diboson backgrounds, a \(20\%\) uncertainty is used to account for modelling uncertainties in these processes~\cite{ATL-PHYS-PUB-2016-005}.
For the dark Higgs signal model, uncertainties in the renormalisation and factorisation scale choices are evaluated following the same procedure as used for SM backgrounds.
The uncertainty in the luminosity is also included and has a negligible impact.


%
\section{Statistical interpretation}
\label{sec:stats}

%
The statistical interpretation of the search results is performed via simultaneous likelihood fits to the observed data, using the \texttt{HistFitter} framework~\cite{Baak:2014wma}.
The likelihood consists of a product of Poisson probability density functions describing the observed and expected numbers of events in the relevant regions.
The systematic uncertainties in the backgrounds and signal processes are included as Gaussian distributed nuisance parameters in the likelihood expression.
For the background processes scaled by normalisation factors (\ttbarl, \ttbarb, \Zjets), these appear as freely floating parameters multiplying the expected MC yields.
MC statistical uncertainties are included as nuisance parameters in the likelihood with Poisson distributions.
A background-only fit is first performed to validate the SM background estimate.
For this fit, only the four control region yields are included, and it is assumed that no signal is present.
The fit constrains the values of the background normalisation factors, and the results of the fit are applied to the validation region to assess the post-fit agreement between the observed \VR\ data and background prediction.

To test the SM-only hypothesis and place limits on the possible number of BSM signal events entering each \DR\ at \(95\%\) confidence level (CL) without making explicit assumptions about the signal \mhh\ shape, fits are performed.
One fit is performed for each \DR\ in turn, simultaneously fitting the four control regions and the \DR\@.
A dummy signal is included in each \DR, whose signal strength is the parameter of interest for the fit. It is assumed the signal contributions to the control regions are negligible.

To place limits on the dark Higgs models at \(95\%\) CL using the \(CL_s\) method~\cite{Read:2002hq}, the four control regions and the binned \SR\ are fitted simultaneously.
The parameter of interest in the fit is the signal strength \(\mu_\text{sig}\), which multiplies the signal yield nominally predicted by the model.


%
\section{Results}
\label{sec:results}

%

%
The background normalisation factors obtained in the background-only fit are: \(\mu_{\Zjets} = 1.24 \pm 0.14\), \(\mu_{\ttbarb} = 1.2 \pm 0.6\) and \(\mu_{\ttbarl} = 0.91 \pm 0.14\).  %
The agreement between the observed data and the SM background prediction after the background-only fit is shown in Figure~\ref{fig:CRVRpostfitSummary}.
The \mhh\ distribution in the VR after the background-only fit is shown in Figure~\ref{fig:postfitmhhVR}, demonstrating no mismodelling trend and a small upward fluctuation (\(<1.7\sigma\)).
The distribution of the NN score after the background-only fit is shown in Figure~\ref{fig:postfitkin:score}, for the range of score values used in the analysis regions.
A good agreement between the SM background prediction and data is observed throughout the range.
The \mhh\ distribution in the binned signal region after the background-only fit is shown in Figure~\ref{fig:postfitkin:mhh}.
The overlaid benchmark signal points highlight the good discrimination power of the NN and \mhh\ distributions.
There are no significant excesses or deficits in the data compared with the SM prediction.
Additionally, the distribution has a similar shape to that seen in the VR in Figure~\ref{fig:postfitmhhVR}, demonstrating that the NN score is not highly correlated with \mhh, in line with the NN design.

\begin{figure}[t]
\centering
\includegraphics[width=0.6\textwidth]{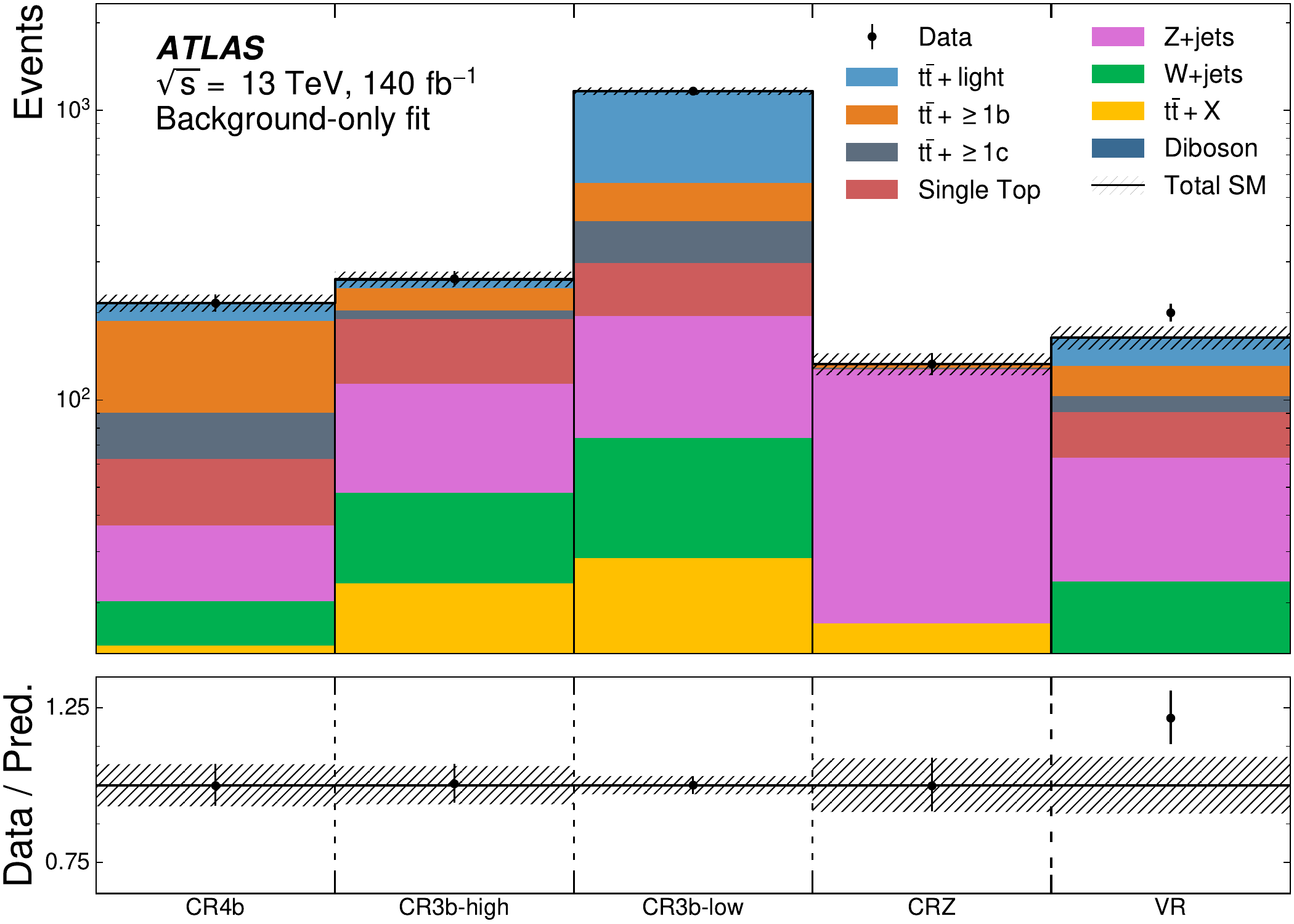}
\caption{Observed data and expected background yields in each control and validation region after the background-only fit. The hashed band shows the total systematic and MC statistical uncertainty. In the lower panel the ratio of data and background yields is shown.}
\label{fig:CRVRpostfitSummary}
\end{figure}

\begin{figure}[t]
\centering
\includegraphics[width=0.6\textwidth]{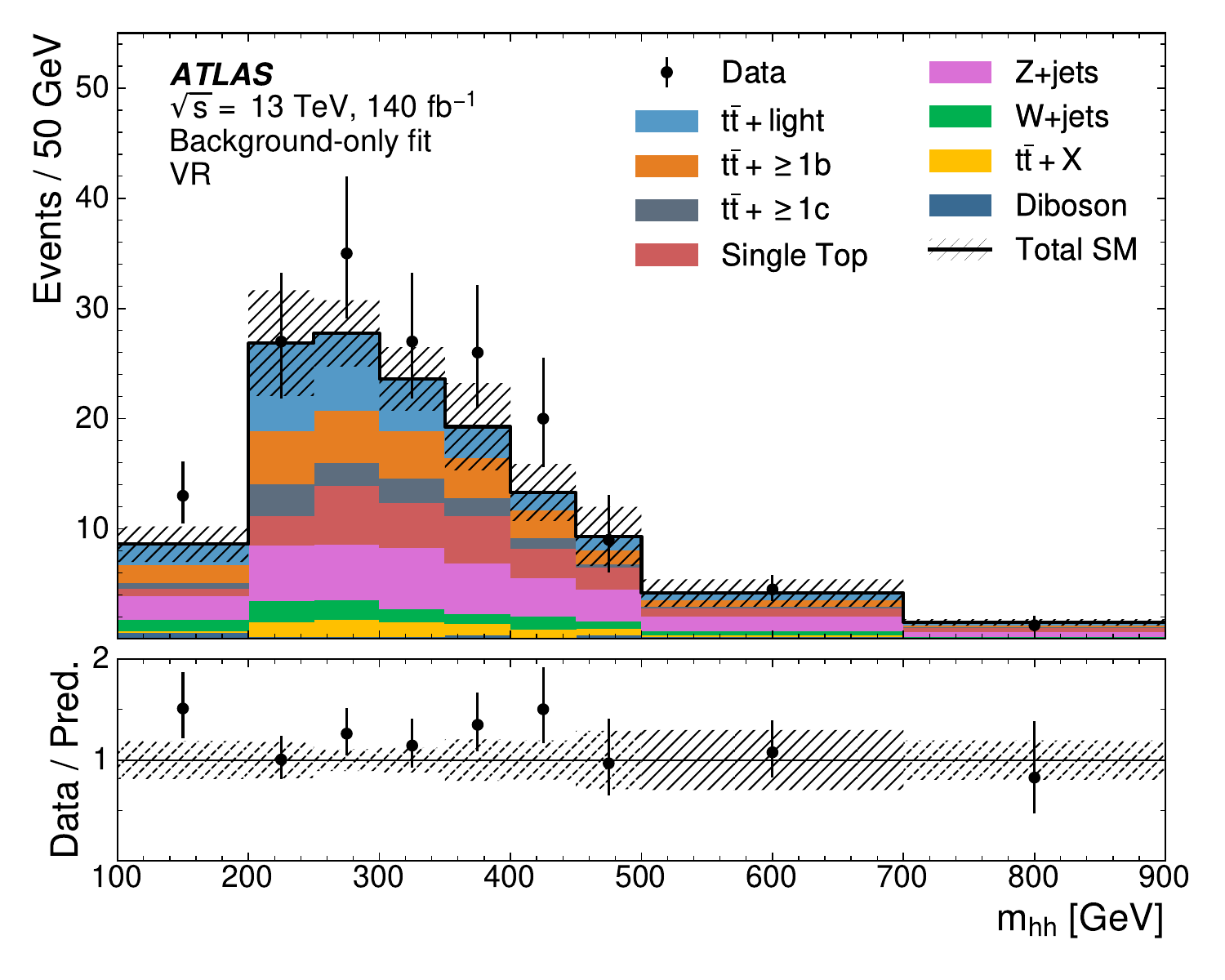}
\caption{The observed data and expected background \mhh\ distribution in the \VR, after the background-only fit. The hashed bands show the total systematic and MC statistical uncertainty. The lower panel shows the ratio of data and predicted background yields.}
\label{fig:postfitmhhVR}
\end{figure}

\begin{figure}[t]
\centering
\subfloat[]{\label{fig:postfitkin:score}\includegraphics[width=0.49\textwidth]{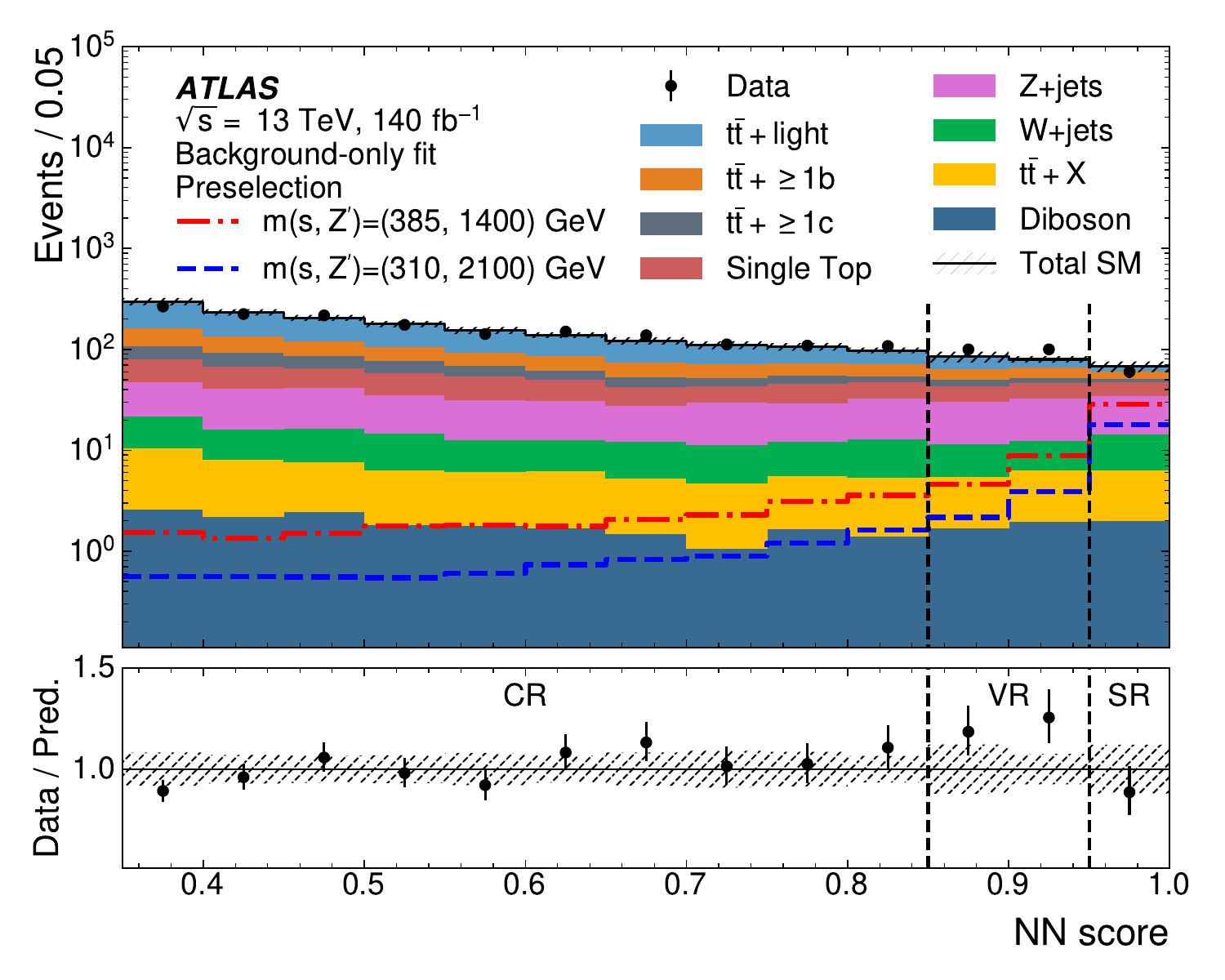}}
\subfloat[]{\label{fig:postfitkin:mhh}\includegraphics[width=0.49\textwidth]{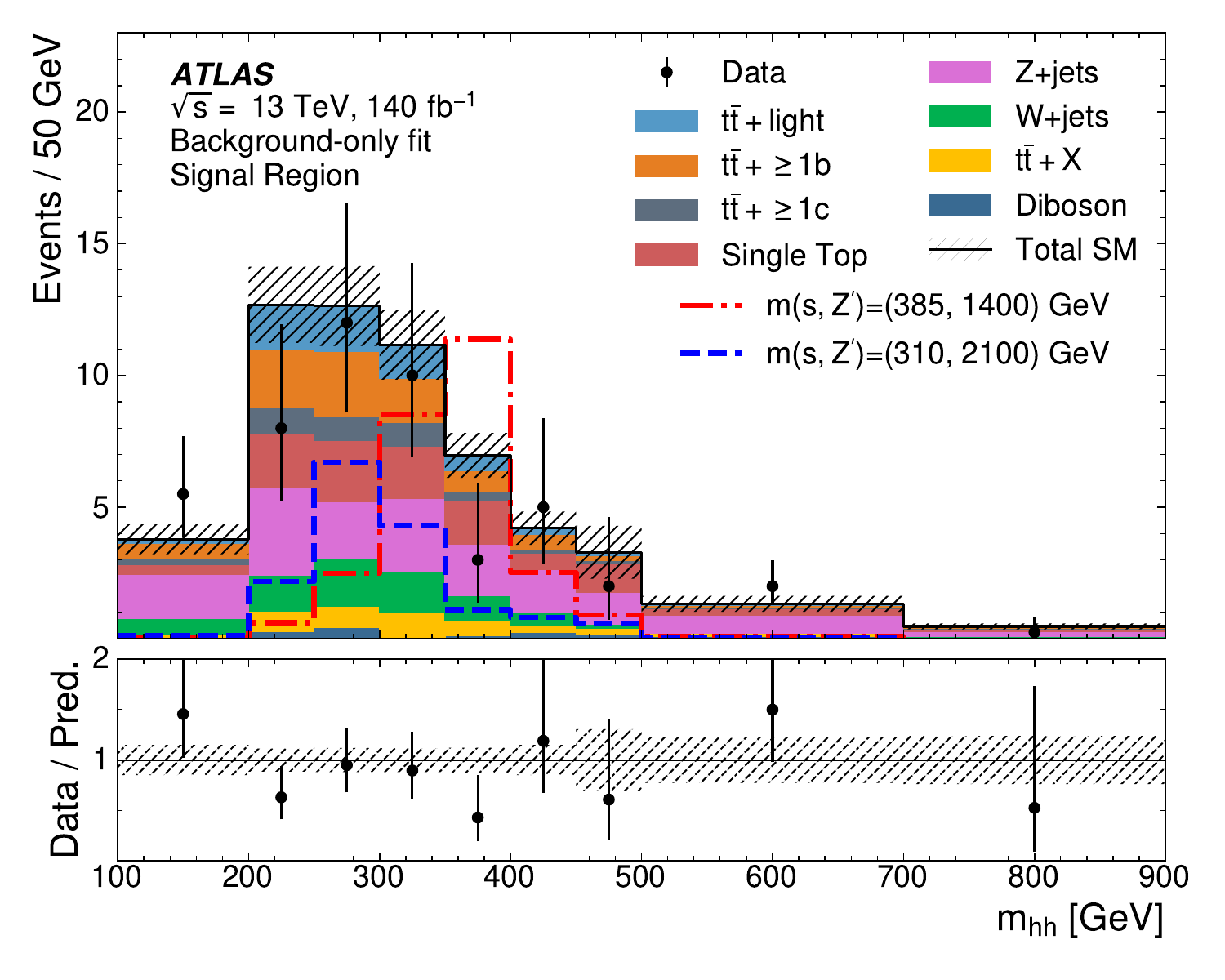}}
\caption{The observed data and expected background (a) NN score distribution after the preselection requirements and (b) \mhh\ distribution in the \SR, after the background-only fit. The hashed bands show the total systematic and MC statistical uncertainty. Benchmark signal distributions are overlaid as dashed lines. The lower panel shows the ratio of data and predicted background yields.}
\label{fig:postfitkin}
\end{figure}

To place limits on the possible number of BSM events of a similar topology entering each \DR, more model-agnostic fits are performed, as described in Section~\ref{sec:stats}.
The post-fit agreement between the observed data and the expected SM background yields follows from the fitted \mhh\ distribution, and is shown in Figure~\ref{fig:disco}, with the derived upper limits detailed in Table~\ref{tab:discolim}.

\begin{figure}[t]
\centering
\includegraphics[width=0.9\textwidth]{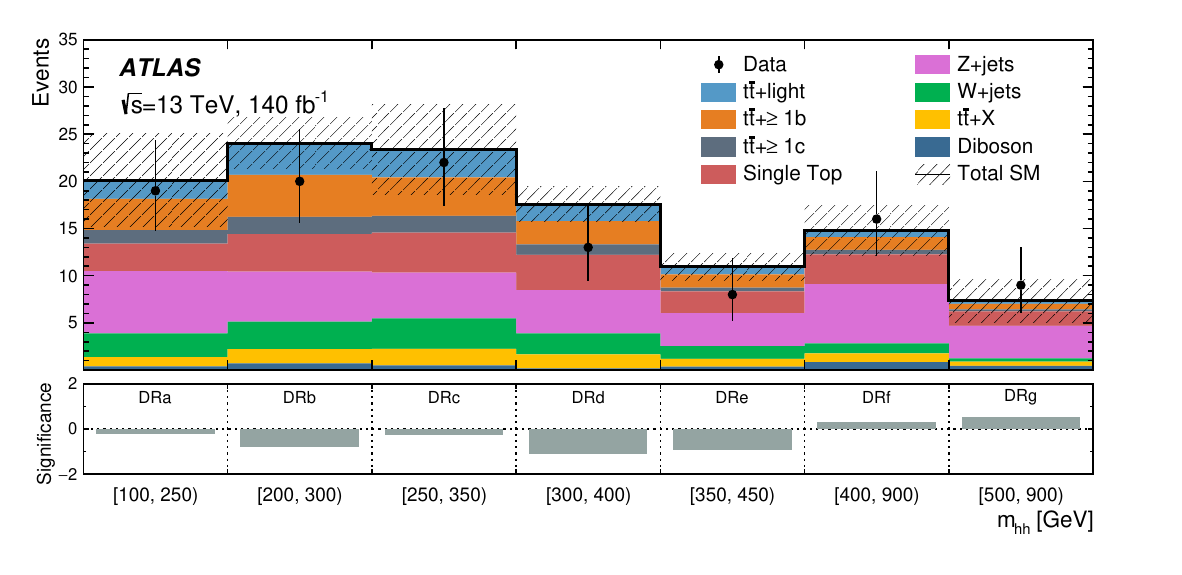}
\caption{Observed data and fitted background yields in each \DR\ region. The hashed band shows the total systematic and MC statistical uncertainty. In the lower panel the statistical significance (measured using Eq. 1 of Ref.~\cite{ATL-PHYS-PUB-2020-025}) of fluctuations between the data and SM expectation is shown.}
\label{fig:disco}
\end{figure}

\begin{table}[t]
\centering
\caption{Upper limits on the expected (observed) number of BSM events \(S^{95}_{\text{exp}}\) (\(S^{95}_{\text{obs}}\)) at 95\% confidence level, the \(\pm1\sigma\) variations on the expected upper limit, and the observed limit on the effective visible BSM cross-section (\(\langle\epsilon \sigma^{95}_{\text{obs}}\rangle\)), for each of \DRa-\DRg. The discovery \(p_0\)-value is also shown, capped at 0.5 if the data yield is smaller than the SM prediction.
}
\begin{tabular}{|lc|cccccc|}
\hline
Region  & \mhh \([\GeV]\)   & Data & Fitted background & \(S^{95}_{\text{exp}}\)      & \(S^{95}_{\text{obs}}\)  & \(\langle\epsilon \sigma^{95}_{\text{obs}}\rangle\) [fb] & \(p_0\)-value \\[1ex]
\hline
\DRa    & \([100,250)\)     & 19 & $20.1\pm 4.9$ & $11.1_{-3.3}^{+5.1}$  & 10                & 0.072                                             & 0.50 \\
\DRb    & \([200,300)\)     & 20 & $24.1\pm 2.7$ & $12.1_{-3.6}^{+5.5}$  & 8.8               & 0.063                                             & 0.50 \\
\DRc    & \([250,350)\)     & 22 & $23.4\pm 4.7$ & $11.9_{-3.6}^{+5.5}$  & 11                & 0.076                                             & 0.50 \\
\DRd    & \([300,400)\)     & 13 & $17.6\pm 1.9$ & $10.2_{-3.1}^{+4.9}$  & 6.9               & 0.049                                             & 0.50 \\
\DRe    & \([350,450)\)     & 8  & $10.9\pm 1.4$ & $8.2_{-2.5}^{+4.1}$   & 5.9               & 0.042                                             & 0.50 \\
\DRf    & \([400,900)\)     & 16 & $14.8\pm 2.7$ & $10.0_{-3.0}^{+4.7}$  & 11                & 0.079                                             & 0.40 \\
\DRg    & \([500,900)\)     & 9  & $~~7.4\pm 2.3$ & $7.2_{-2.3}^{+3.6}$  & 8.7               & 0.062                                             & 0.33 \\
\hline
\end{tabular}
\label{tab:discolim}
\end{table}

In the absence of evidence for a signal, exclusion limits are set on the dark Higgs model in the plane of the masses of the dark Higgs boson (\(s\)) and \(Z'\) mediator.
The impacts of each category of uncertainties in the \SR\ yields in the exclusion fits are assessed in Table~\ref{tab:systs} for a representative signal point with \(\mZprime =\SI{1400}{\GeV}\) and \(\ms = \SI{385}{\GeV}\).
Consistent with the limited yields in the SR, the results are statistically limited, with the statistical uncertainty of the data dominating. The larger systematic uncertainty components are those from the modelling of the predicted backgrounds, such as the normalisation of the freely floating components and the theoretical modelling and cross-section uncertainties of the other sub-leading components. Smaller contributions come from the limited size of the Monte-Carlo simulations and experimental uncertainties pertaining to the jet and \(b\)-jet calibration and identification.

%
\begin{table}[!h]
\centering
\caption{Summary of the relative impact of each category of uncertainty in the \SR\ yield, in a signal plus background fit to a \(\mZprime =\SI{1400}{\GeV}\), \(\ms = \SI{385}{\GeV}\) signal simulation.}
\label{tab:systs}
\begin{tabular}{|l|c|}
\hline
Source of uncertainty & Percentage of total uncertainty [\%] \\
\hline
Signal modelling & <1 \\
\(b\)-jet identification & 5 \\
Jet calibrations & 9 \\
Other experimental & 2 \\
\ttbar\ modelling \& normalisation & 11 \\
\Zjets modelling \& normalisation & 9 \\
Single top modelling \& cross-section & 18  \\
Other background modelling \& cross-section & 13 \\
MC statistical uncertainty & 17 \\
\hline
Overall systematic error & 33 \\
Data statistical & 95 \\
Total uncertainty & 100 \\
\hline
\end{tabular}
\end{table}


The observed and expected exclusion limits are calculated from the binned signal region, and shown in Figure~\ref{fig:limit}.
The observed limits follow the expectation closely, and are constrained by the large acceptance changes at the \ZpDM threshold of \(\mZprime \sim \SI{400}{\GeV}\) and the \(s\)-decay thresholds at \(\ms \sim \SI{250}{\GeV}\) and \(\SI{400}{\GeV}\).
At high \mZprime, the falling cross-section leads to a wider variation in the expected limits, and small fluctuations in the data result in slightly stronger or weaker exclusions than the median expectation accordingly.
A comparison of the observed contours with limits placed from other ATLAS dark Higgs boson searches~\cite{EXOT-2020-26,EXOT-2020-04,EXOT-2018-40} is shown in Figure~\ref{fig:limitsummary}, highlighting the complementary sensitivities that different dark Higgs boson signatures have to different areas of mass parameter space.
The di-Higgs decay of the dark Higgs boson is sensitive to \(\ms > \SI{250}{\GeV}\) and the high Higgs boson branching ratio to \(bb\) provides unique sensitivity in this region, where other dibosonic final states lose sensitivity from their decreased branching fraction and the increasing collimation of the decay products. Also shown in Figure~\ref{fig:limitsummary} is the predicted DM relic density as calculated by MadDM~\cite{Ambrogi:2018jqj}. At \(\ms \simeq 2 \mchi = 400\)~\GeV, DM annihilation via a dark Higgs boson is enhanced, strongly depleting the surviving relic abundance of DM\@.
For this benchmark with \(\mchi = \SI{200}{\GeV}\), the \(s \rightarrow \chi\chi \) decay dominates above this \ms\ value; however these results have generic sensitivity to \(\ms > \SI{400}{\GeV}\), e.g. in scenarios with larger \mchi.

\begin{figure}[t]
\centering
\includegraphics[width=0.8\textwidth]{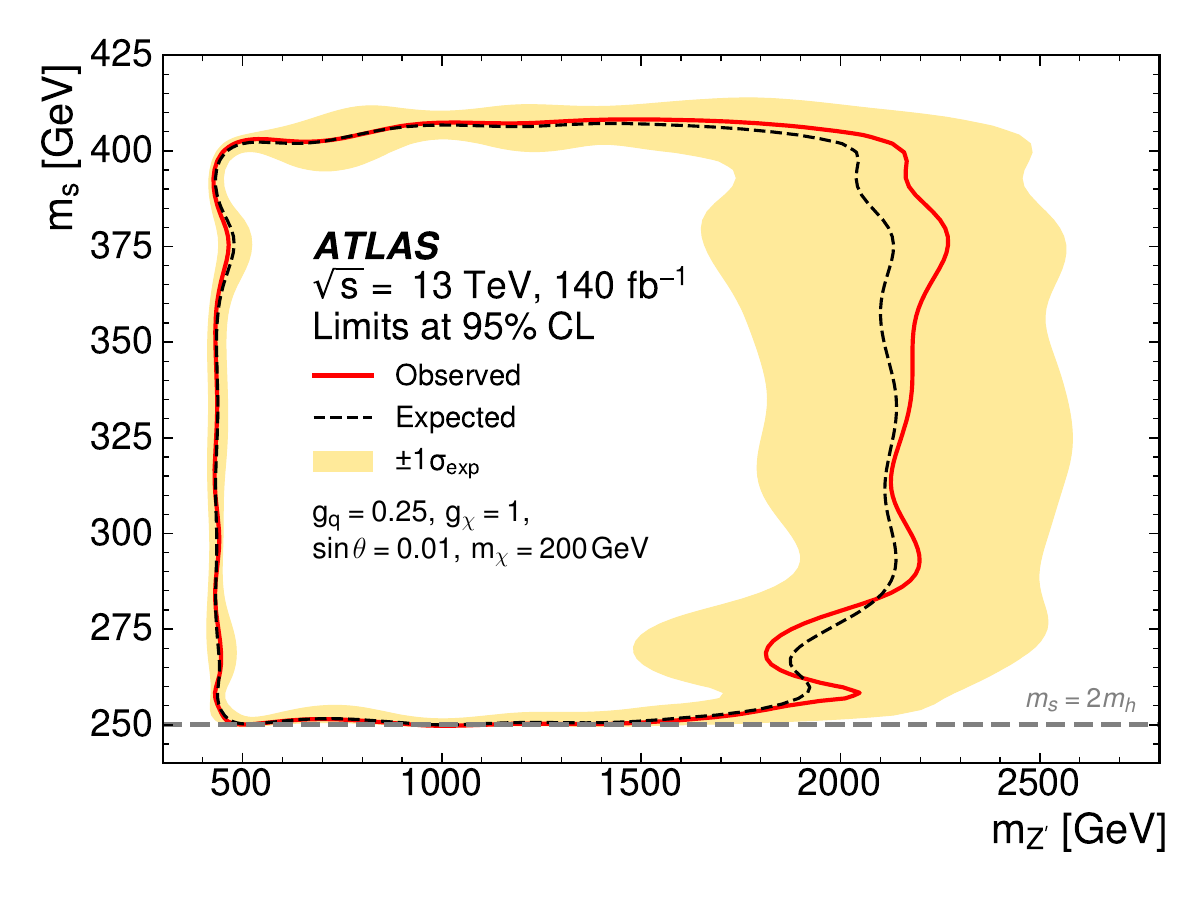}
\caption{Observed and expected exclusion limits in the \((\mZprime,\ms)\) plane at \(95\%\) CL, with the excluded part inside the closed contour. The \(\pm1\sigma\) expected exclusion intervals are shown as the filled bands. The horizontal dashed line at \(\ms=2\mh\) indicates the limit where the \(s\rightarrow h h\) decay mode becomes kinematically accessible.}
\label{fig:limit}
\end{figure}

\begin{figure}[t]
\centering
\includegraphics[width=0.9\textwidth]{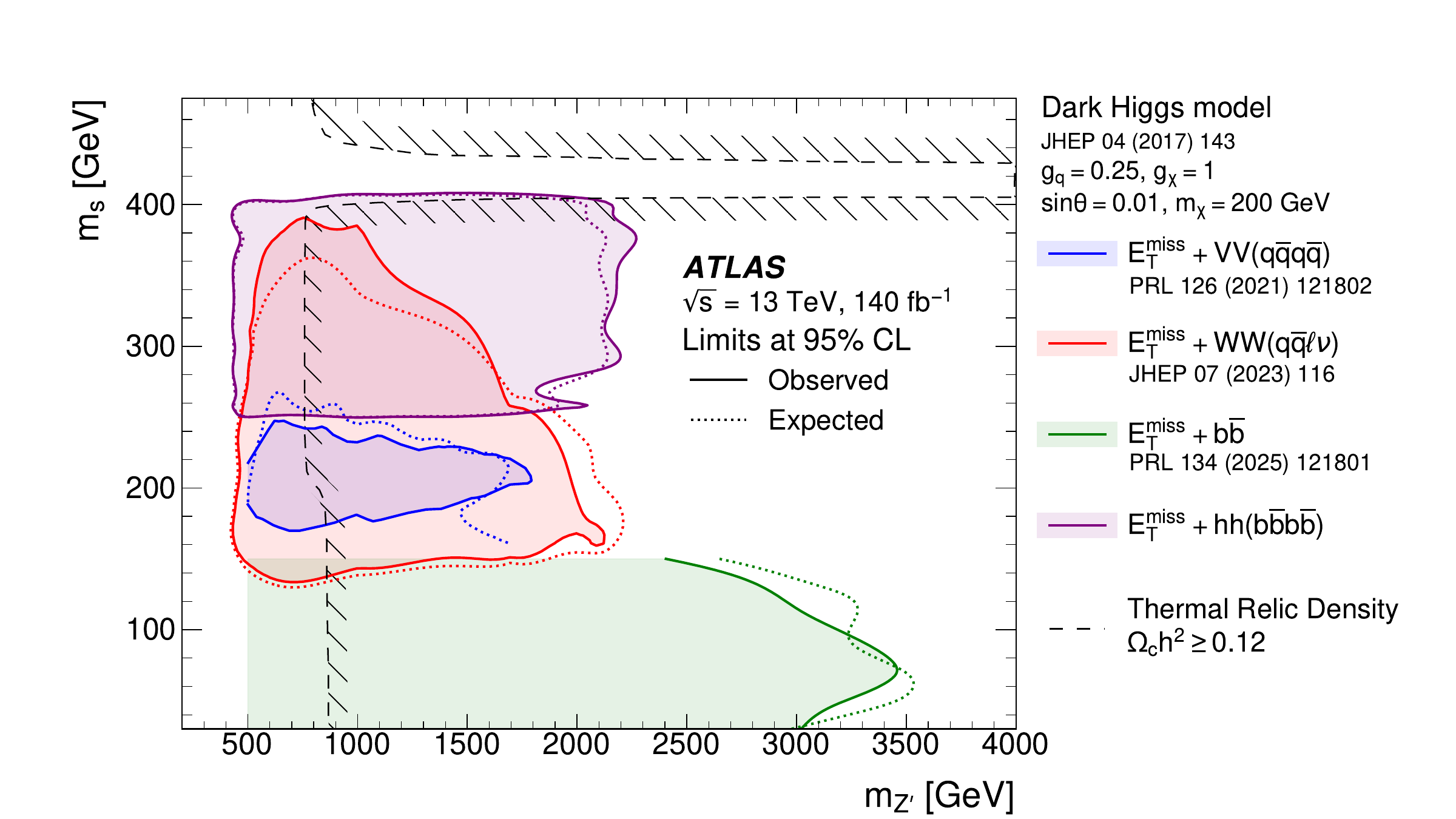}
\caption{A comparison of this analysis' exclusions with existent ATLAS searches~\cite{EXOT-2018-40, EXOT-2020-04, EXOT-2020-26}. Their observed and expected exclusion limits in the \((\mZprime,\ms)\) plane at \(95\%\) CL are shown. The observed relic density is indicated with a dashed line, with the diagonal lines indicating an overabundance of DM.}
\label{fig:limitsummary}
\end{figure}


\FloatBarrier

\section{Conclusion}
\label{sec:conclusion}

A search for dark matter in a final state with significant \ptmiss\ and a resonantly produced pair of Higgs bosons resulting in at least three \(b\)-tagged jets is performed, using the \(\SI{140}{\ifb}\) of Run 2 data recorded by the ATLAS detector at the LHC between 2015 and 2018.
The observed data are found to be consistent with Standard Model predictions.
Upper limits are placed at 95\% confidence level on the effective production cross-section for new particle states with mass greater than \(\SI{250}{\GeV}\) decaying into \(hh\) in association with \ptmiss.
The result is interpreted using a dark Higgs model, with a dark Higgs boson \(s\) giving rise to massive dark matter, and a \(Z'\) mediator.
Upper limits are derived on the hypothesised particle masses for \(250 < \ms < \SI{400}{\GeV}\), excluding \(Z'\) masses up to \(\SI{2.3}{\TeV}\) in a recommended model benchmark.
This new signature significantly extends the constraints in this mass region existing from previous dark Higgs analyses by the ATLAS and CMS collaborations in different final states.
The results complement other lower-mass dark Higgs boson searches, searches for extended Higgs sectors and collider DM searches more generally.


%
%
%
%
%
\section*{Acknowledgements}
%

%

%
%

%
%

We thank CERN for the very successful operation of the LHC and its injectors, as well as the support staff at
CERN and at our institutions worldwide without whom ATLAS could not be operated efficiently.

The crucial computing support from all WLCG partners is acknowledged gratefully, in particular from CERN, the ATLAS Tier-1 facilities at TRIUMF/SFU (Canada), NDGF (Denmark, Norway, Sweden), CC-IN2P3 (France), KIT/GridKA (Germany), INFN-CNAF (Italy), NL-T1 (Netherlands), PIC (Spain), RAL (UK) and BNL (USA), the Tier-2 facilities worldwide and large non-WLCG resource providers. Major contributors of computing resources are listed in Ref.~\cite{ATL-SOFT-PUB-2025-001}.

We gratefully acknowledge the support of ANPCyT, Argentina; YerPhI, Armenia; ARC, Australia; BMWFW and FWF, Austria; ANAS, Azerbaijan; CNPq and FAPESP, Brazil; NSERC, NRC and CFI, Canada; CERN; ANID, Chile; CAS, MOST and NSFC, China; Minciencias, Colombia; MEYS CR, Czech Republic; DNRF and DNSRC, Denmark; IN2P3-CNRS and CEA-DRF/IRFU, France; SRNSFG, Georgia; BMFTR, HGF and MPG, Germany; GSRI, Greece; RGC and Hong Kong SAR, China; ICHEP and Academy of Sciences and Humanities, Israel; INFN, Italy; MEXT and JSPS, Japan; CNRST, Morocco; NWO, Netherlands; RCN, Norway; MNiSW, Poland; FCT, Portugal; MNE/IFA, Romania; MSTDI, Serbia; MSSR, Slovakia; ARIS and MVZI, Slovenia; DSI/NRF, South Africa; MICIU/AEI, Spain; SRC and Wallenberg Foundation, Sweden; SERI, SNSF and Cantons of Bern and Geneva, Switzerland; NSTC, Taipei; TENMAK, T\"urkiye; STFC/UKRI, United Kingdom; DOE and NSF, United States of America.

Individual groups and members have received support from BCKDF, CANARIE, CRC and DRAC, Canada; CERN-CZ, FORTE and PRIMUS, Czech Republic; COST, ERC, ERDF, Horizon 2020, ICSC-NextGenerationEU and Marie Sk{\l}odowska-Curie Actions, European Union; Investissements d'Avenir Labex, Investissements d'Avenir Idex and ANR, France; DFG and AvH Foundation, Germany; Herakleitos, Thales and Aristeia programmes co-financed by EU-ESF and the Greek NSRF, Greece; BSF-NSF and MINERVA, Israel; NCN and NAWA, Poland; La Caixa Banking Foundation, CERCA Programme Generalitat de Catalunya and PROMETEO and GenT Programmes Generalitat Valenciana, Spain; G\"{o}ran Gustafssons Stiftelse, Sweden; The Royal Society and Leverhulme Trust, United Kingdom.

In addition, individual members wish to acknowledge support from CERN: European Organization for Nuclear Research (CERN DOCT); Chile: Agencia Nacional de Investigaci\'on y Desarrollo (FONDECYT 1230812, FONDECYT 1240864); China: Chinese Ministry of Science and Technology (MOST-2023YFA1605700, MOST-2023YFA1609300), National Natural Science Foundation of China (NSFC - 12175119, NSFC 12275265); Czech Republic: Czech Science Foundation (GACR - 24-11373S), Ministry of Education Youth and Sports (ERC-CZ-LL2327, FORTE CZ.02.01.01/00/22\_008/0004632), PRIMUS Research Programme (PRIMUS/21/SCI/017); EU: H2020 European Research Council (ERC - 101002463); European Union: European Research Council (BARD No. 101116429, ERC - 948254, ERC 101089007), European Regional Development Fund (SMASH COFUND 101081355, SLO ERDF), Horizon 2020 Framework Programme (MUCCA - CHIST-ERA-19-XAI-00), European Union, Future Artificial Intelligence Research (FAIR-NextGenerationEU PE00000013), Italian Center for High Performance Computing, Big Data and Quantum Computing (ICSC, NextGenerationEU); France: Agence Nationale de la Recherche (ANR-21-CE31-0022, ANR-22-EDIR-0002); Germany: Baden-Württemberg Stiftung (BW Stiftung-Postdoc Eliteprogramme), Deutsche Forschungsgemeinschaft (DFG - 469666862, DFG - CR 312/5-2); China: Research Grants Council (GRF); Italy: Istituto Nazionale di Fisica Nucleare (ICSC, NextGenerationEU), Ministero dell'Università e della Ricerca (NextGenEU I53D23000820006 M4C2.1.1, SOE2024\_0000023); Japan: Japan Society for the Promotion of Science (JSPS KAKENHI JP22H01227, JSPS KAKENHI JP22H04944, JSPS KAKENHI JP22KK0227, JSPS KAKENHI JP24K23939, JSPS KAKENHI JP25H00650, JSPS KAKENHI JP25H01291, JSPS KAKENHI JP25K01023); Norway: Research Council of Norway (RCN-314472); Poland: Ministry of Science and Higher Education (IDUB AGH, POB8, D4 no 9722), Polish National Science Centre (NCN 2021/42/E/ST2/00350, NCN OPUS 2023/51/B/ST2/02507, NCN OPUS nr 2022/47/B/ST2/03059, NCN UMO-2019/34/E/ST2/00393, UMO-2022/47/O/ST2/00148, UMO-2023/49/B/ST2/04085, UMO-2023/51/B/ST2/00920, UMO-2024/53/N/ST2/00869); Portugal: Foundation for Science and Technology (FCT); Spain: Ministry of Science and Innovation (MCIN \& NextGenEU PCI2022-135018-2, MICIN \& FEDER PID2021-125273NB, RYC2019-028510-I, RYC2020-030254-I, RYC2021-031273-I, RYC2022-038164-I); Sweden: Carl Trygger Foundation (Carl Trygger Foundation CTS 22:2312), Swedish Research Council (Swedish Research Council 2023-04654, VR 2021-03651, VR 2022-03845, VR 2022-04683, VR 2023-03403, VR 2024-05451), Knut and Alice Wallenberg Foundation (KAW 2018.0458, KAW 2022.0358, KAW 2023.0366); Switzerland: Swiss National Science Foundation (SNSF - PCEFP2\_194658); United Kingdom: Leverhulme Trust (Leverhulme Trust RPG-2020-004), Royal Society (NIF-R1-231091); United States of America: U.S. Department of Energy (ECA DE-AC02-76SF00515), Neubauer Family Foundation.

%
%


%
\clearpage
\printbibliography
\clearpage
\input{atlas_authlist}
%

%
%

%

%
%
%
%

\end{document}

%% file: atlas_authlist.tex
 
\begin{flushleft}
\hypersetup{urlcolor=black}
{\Large The ATLAS Collaboration}

\bigskip

\AtlasOrcid[0000-0002-6665-4934]{G.~Aad}$^\textrm{\scriptsize 104}$,
\AtlasOrcid[0000-0001-7616-1554]{E.~Aakvaag}$^\textrm{\scriptsize 17}$,
\AtlasOrcid[0000-0002-5888-2734]{B.~Abbott}$^\textrm{\scriptsize 123}$,
\AtlasOrcid[0000-0002-0287-5869]{S.~Abdelhameed}$^\textrm{\scriptsize 119a}$,
\AtlasOrcid[0000-0002-1002-1652]{K.~Abeling}$^\textrm{\scriptsize 55}$,
\AtlasOrcid[0000-0001-5763-2760]{N.J.~Abicht}$^\textrm{\scriptsize 49}$,
\AtlasOrcid[0000-0002-8496-9294]{S.H.~Abidi}$^\textrm{\scriptsize 30}$,
\AtlasOrcid[0009-0003-6578-220X]{M.~Aboelela}$^\textrm{\scriptsize 45}$,
\AtlasOrcid[0000-0002-9987-2292]{A.~Aboulhorma}$^\textrm{\scriptsize 36e}$,
\AtlasOrcid[0000-0001-5329-6640]{H.~Abramowicz}$^\textrm{\scriptsize 157}$
% ...
\bigskip
\\

% ...
$^{17}$Department for Physics and Technology, University of Bergen, Bergen; Norway.\\
% ...
$^{30}$Physics Department, Brookhaven National Laboratory, Upton NY; United States of America.\\
% ...
$^{36}$$^{(a)}$Facult\'e des Sciences Ain Chock, Universit\'e Hassan II de Casablanca;$^{(b)}$Facult\'{e} des Sciences, Universit\'{e} Ibn-Tofail, K\'{e}nitra;$^{(c)}$Facult\'e des Sciences Semlalia, Universit\'e Cadi Ayyad, LPHEA-Marrakech;$^{(d)}$LPMR, Facult\'e des Sciences, Universit\'e Mohamed Premier, Oujda;$^{(e)}$Facult\'e des sciences, Universit\'e Mohammed V, Rabat;$^{(f)}$Institute of Applied Physics, Mohammed VI Polytechnic University, Ben Guerir; Morocco.\\
% ...
$^{45}$Physics Department, Southern Methodist University, Dallas TX; United States of America.\\
% ...
$^{49}$Fakult\"{a}t Physik , Technische Universit{\"a}t Dortmund, Dortmund; Germany.\\
% ...
$^{55}$II. Physikalisches Institut, Georg-August-Universit\"{a}t G\"ottingen, G\"ottingen; Germany.\\
% ...
$^{104}$CPPM, Aix-Marseille Universit\'e, CNRS/IN2P3, Marseille; France.\\
% ...
$^{119}$$^{(a)}$New York University Abu Dhabi, Abu Dhabi;$^{(b)}$United Arab Emirates University, Al Ain; United Arab Emirates.\\
% ...
$^{123}$Homer L. Dodge Department of Physics and Astronomy, University of Oklahoma, Norman OK; United States of America.\\
% ...
$^{157}$Raymond and Beverly Sackler School of Physics and Astronomy, Tel Aviv University, Tel Aviv; Israel.\\
% ...
$^{a}$ Also at Affiliated with an institute formerly covered by a cooperation agreement with CERN.\\
$^{b}$ Also at An-Najah National University, Nablus; Palestine.\\
$^{c}$ Also at Borough of Manhattan Community College, City University of New York, New York NY; United States of America.\\
$^{d}$ Also at Center for Interdisciplinary Research and Innovation (CIRI-AUTH), Thessaloniki; Greece.\\
$^{e}$ Also at Centre of Physics of the Universities of Minho and Porto (CF-UM-UP); Portugal.\\
$^{f}$ Also at CERN, Geneva; Switzerland.\\
% ...

\end{flushleft}
